\documentclass[superscriptaddress,twocolumn,10pt,prx,aps,showpacs,longbibliography,floatfix]{revtex4-2}

\usepackage[dvipsnames]{xcolor}

\definecolor{darkred}{rgb}{0.6,0.05,0.05}
\definecolor{darkgreen}{rgb}{0.05,0.6,0.05}
\definecolor{darkblue}{rgb}{0.05,0.05,0.6}
\usepackage[colorlinks]{hyperref}
\hypersetup{colorlinks,
linkcolor=darkred,
filecolor=darkgreen,
urlcolor=darkblue,
citecolor=darkgreen}

\usepackage{textcomp,mathcomp}
\usepackage{mathtools}
\usepackage{amsmath}
\usepackage{amssymb}
\usepackage[normalem]{ulem}
\usepackage{amsthm}
\usepackage{xfrac}
\usepackage{dsfont}
\usepackage{makecell}
\usepackage{physics}
\usepackage{graphicx}
\usepackage{hyperref}
\usepackage{dcolumn}
\usepackage{bm}

\usepackage{multirow}
\usepackage{tabularx}
\usepackage{booktabs}

\usepackage{cleveref}
\crefname{equation}{Eq.}{Eqs.}
\Crefname{equation}{Equation}{Equations}
\crefname{figure}{Fig.}{Figs.}
\Crefname{figure}{Figure}{Figures}
\crefname{section}{Sec.}{Sects.}
\Crefname{section}{Section}{Sections}
\crefname{table}{Table}{Tables}
\crefname{appendix}{Appendix}{Apps.}
\Crefname{appendix}{Appendix}{Apps.}

\usepackage{soul}

\definecolor{armygreen}{rgb}{0.29, 0.33, 0.13}
\definecolor{palatinatepurple}{rgb}{0.41, 0.16, 0.38}
\definecolor{sangria}{rgb}{0.57, 0.0, 0.04}

\makeatletter
\newcommand*\bigcdot{\mathpalette\bigcdot@{.5}}
\newcommand*\bigcdot@[2]{\mathbin{\vcenter{\hbox{\scalebox{#2}{$\m@th#1\bullet$}}}}}
\makeatother

\renewcommand{\paragraph}[1]{\vspace{5pt}\noindent \textbf{#1--.}}

\begin{document}

\author{Guillaume Beaulieu}
\affiliation{Hybrid Quantum Circuits Laboratory (HQC), Institute of Physics, \'{E}cole Polytechnique F\'{e}d\'{e}rale de Lausanne (EPFL), 1015 Lausanne, Switzerland}
\affiliation{Center for Quantum Science and Engineering, \\ \'{E}cole Polytechnique F\'{e}d\'{e}rale de Lausanne (EPFL), CH-1015 Lausanne, Switzerland}
\author{Jun-Zhe Chen}
\affiliation{Hybrid Quantum Circuits Laboratory (HQC), Institute of Physics, \'{E}cole Polytechnique F\'{e}d\'{e}rale de Lausanne (EPFL), 1015 Lausanne, Switzerland}
\affiliation{Center for Quantum Science and Engineering, \\ \'{E}cole Polytechnique F\'{e}d\'{e}rale de Lausanne (EPFL), CH-1015 Lausanne, Switzerland}
\author{Marco Scigliuzzo}
\affiliation{Center for Quantum Science and Engineering, \\ \'{E}cole Polytechnique F\'{e}d\'{e}rale de Lausanne (EPFL), CH-1015 Lausanne, Switzerland}
\affiliation{Laboratory of Photonics and Quantum Measurements (LPQM),  Institute of Physics, EPFL, CH-1015 Lausanne, Switzerland}
\author{Othmane Benhayoune-Khadraoui}
\affiliation{Institut Quantique and D\'{e}partement de Physique, Universit\'{e} de Sherbrooke, Sherbrooke J1K 2R1 Quebec, Canada}
\author{Alex A. Chapple}
\affiliation{Institut Quantique and D\'{e}partement de Physique, Universit\'{e} de Sherbrooke, Sherbrooke J1K 2R1 Quebec, Canada}
\author{Peter A. Spring}
\affiliation{RIKEN Center for Quantum Computing (RQC), Wako, Saitama 351-0198, Japan}
\author{Alexandre Blais}
\affiliation{Institut Quantique and D\'{e}partement de Physique, Universit\'{e} de Sherbrooke, Sherbrooke J1K 2R1 Quebec, Canada}
\affiliation{CIFAR, Toronto, ON M5G 1M1, Canada}

\author{Pasquale Scarlino}
\email[E-mail: ]{pasquale.scarlino@epfl.ch}
\affiliation{Hybrid Quantum Circuits Laboratory (HQC), Institute of Physics, \'{E}cole Polytechnique F\'{e}d\'{e}rale de Lausanne (EPFL), 1015 Lausanne, Switzerland}
\affiliation{Center for Quantum Science and Engineering, \\ \'{E}cole Polytechnique F\'{e}d\'{e}rale de Lausanne (EPFL), CH-1015 Lausanne, Switzerland}

\title{Fast, high-fidelity Transmon readout with intrinsic Purcell protection via nonperturbative cross-Kerr coupling}

\date{\today}

\begin{abstract}

Dispersive readout of superconducting qubits relies on a transverse capacitive coupling that hybridizes the qubit with the readout resonator, subjecting the qubit to Purcell decay and measurement-induced state transitions (MIST). Despite the widespread use of Purcell filters to suppress qubit decay and near-quantum-limited amplifiers, dispersive readout often lags behind single- and two-qubit gates in both speed and fidelity. Here, we experimentally demonstrate \emph{junction readout}, a simple readout architecture that realizes a strong qubit–resonator cross-Kerr interaction without relying on a transverse coupling. This interaction is achieved by coupling a transmon qubit to its readout resonator through both a capacitance and a Josephson junction. By varying the qubit frequency, we show that this hybrid coupling provides intrinsic Purcell protection and enhanced resilience to MIST, enabling readout at high photon numbers. While junction readout is compatible with conventional linear measurement, in this work we exploit the nonlinear coupling to intentionally engineer a large Kerr nonlinearity in the resonator, enabling bifurcation-based readout. Using this approach, we achieve a $99.4$\,\% assignment fidelity with a $68$ ns integration time and a $98.4$\,\% QND fidelity \emph{without} an external Purcell filter or a near-quantum-limited amplifier. These results establish the junction readout architecture with bifurcation-based readout as a scalable and practical alternative to dispersive readout, enabling fast, high-fidelity qubit measurement with reduced hardware overhead.

\end{abstract}

\maketitle

\section{Introduction}
 \label{sec:introduction}

As superconducting quantum processors scale toward fault-tolerant quantum error correction (QEC) \cite{Fowler2012,Marques2021,Krinner2022,Zhao2022,google_2023,Google_2024}, qubit readout has emerged as a critical bottleneck, often lagging behind single- and two-qubit gates in both speed and fidelity \cite{google_2023,Google_2024}. Dispersive readout, the standard method for measuring superconducting qubits, relies on a transverse capacitive coupling between the qubit and the resonator \cite{Blais2004,KochPRA07}. In the dispersive regime, this interaction produces a qubit-state-dependent shift of the resonator frequency \cite{Blais2004,Wallraff2004,Schuster2007}, enabling a quantum non-demolition (QND) measurement in the idealized limit of negligible qubit–resonator hybridization.

Despite its widespread adoption and continuous improvements \cite{Jeffrey2014,McClure2016,Walter2017,Sunada2022,Swiadek2024,Spring2025}, dispersive readout faces intrinsic limitations. The underlying transverse interaction necessarily hybridizes the qubit and the resonator, opening a channel for Purcell-induced relaxation of the qubit \cite{Purcell1995,Houck2008}. This hybridization also leads to additional detrimental effects such as measurement-induced state transitions (MIST) \cite{Sank2016,Lescanne2019,Shillito2022,Cohen2023,Khezri2023,Dumas2024,Hazra2025, Dai2025}.

To mitigate these issues, state-of-the-art implementations rely on additional hardware components: Purcell filters to suppress qubit decay \cite{Reed2010,Jeffrey2014,Sete2015} and near-quantum-limited amplifiers \cite{Mallet2009,Macklin2015,Walter2017}, allowing operation at lower photon numbers to help preserve the QNDness. While effective, these additional components increase system complexity and footprint, require additional calibration, and present challenges for scalability.

One strategy for reducing the readout-chain footprint is to implement intrinsic Purcell filtering, thereby suppressing qubit relaxation without the need for additional filter components. Proposals in this direction either modify the qubit itself, replacing the transmon with symmetry protected circuits \cite{Roy2017,Dassonneville2020,Pfeiffer2024,Hazra2025}, or by modifying the readout circuitry. Examples of the latter include positioning the feedline at a voltage node of the dressed qubit mode inside the resonator \cite{Sunada2022}, coupling the resonator to the feedline at multiple points to interfere decay pathways \cite{Yen2025}, and exploiting stray capacitances in the readout geometry \cite{Bronn2015}. 

Furthermore, strategies for enabling readout at higher photon numbers include modifying the qubit–resonator interaction or exploring parameter regimes beyond the conventional operating range. One such approach is to operate at extremely large detunings, where the qubit frequency is only a small fraction of the resonator frequency ($\omega_r/\omega_q \sim 10$) \cite{Kurilovich(2025),Conolly2025}. This regime, however, presents its own set of challenges. Another method is to replace the transverse coupling with nonlinear interactions which are predicted to suppress MIST even at large photon numbers \cite{Chapple2025_ionization,Mori_2025_supression}. Such couplings include the longitudinal coupling \cite{Didier2015,Richer_2016} and nonperturbative cross-Kerr interactions \cite{Diniz2013,Dassonneville2020,Ye2024,Pfeiffer2024,Mori2025,Salunkhe_2025,Wang2025}. However, these couplings typically require circuit symmetries or the introduction of ancillary modes. In several of these implementations, the nonlinear interaction simultaneously induces a self-Kerr nonlinearity on the resonator, enabling bifurcation-based readout \cite{Siddiqi2006,Boulant2007,Mallet2009,Schmitt2014} at high photon numbers \cite{Dassonneville2023,Wang2025}.

In this work, we realize the recently proposed \emph{junction readout} architecture \cite{Chapple2025, Wang2025}. Unlike most nonlinear-coupling approaches, junction readout achieves nonperturbative cross-Kerr interaction with a remarkably simple modification of the standard dispersive readout circuit: by adding a Josephson junction in parallel with the conventional capacitive coupling. We demonstrate that this circuit provides intrinsic Purcell protection and preserves QNDness at high readout photon numbers. Leveraging the resonator self-Kerr induced by the coupling junction, we implement a bifurcation-based readout and obtain a readout fidelity of $99.4$\,\% with a $68$ ns integration time and a $98.4$\,\% QND fidelity \emph{without} near-quantum-limited amplification or a dedicated Purcell filter. We note that although this work employs bifurcation-based readout, the junction-readout circuit is equally compatible with conventional linear readout by reducing the resonator impedance to suppress its Kerr nonlinearity; we leave such implementations to future device generations. Compared with the closest experimental realization \cite{Wang2025}, our device features a larger cross-Kerr interaction, operates with fewer photons, and achieves faster readout. Importantly, the ability to selectively tune the transmon frequency enables an in-depth investigation of the device physics, providing direct confirmation of intrinsic Purcell filtering and a quantitative validation of the underlying theoretical model. These results establish junction readout as a promising and scalable architecture that addresses the main limitations of dispersive readout with minimal hardware modifications.

\section{Nonperturbative Cross-Kerr Coupling} 

\begin{figure}
    \centering
    \includegraphics[width=1 \columnwidth]{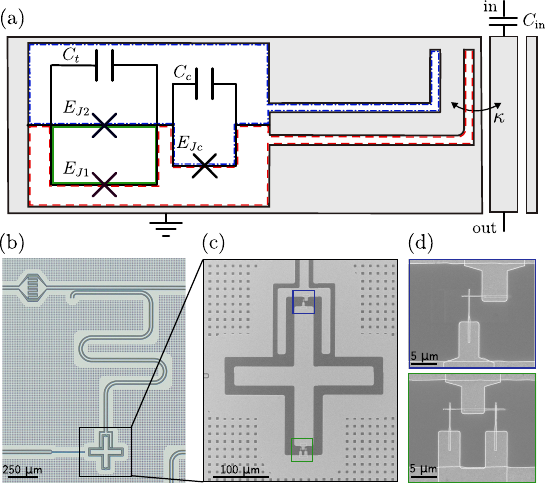}
    \caption{ \textbf{Circuit implementation of junction readout}. (a) Schematic of a single qubit–resonator pair on the device. The circuit contains three flux loops: two large loops that enclose both the qubit and the resonator, highlighted by the red dashed and blue dash–dotted outlines, and a smaller loop formed by the asymmetric SQUID of the transmon, shown in green. (b) Optical micrograph of the first qubit-resonator pair. (c) Scanning electron microscope (SEM) image of the transmon qubit and its coupling to the resonator. (d) SEM images of the coupling junction (highlighted in blue) and the asymmetric SQUID of the transmon (highlighted in green).} 
    \label{Fig:device Schematic}
\end{figure}

Junction readout is implemented by coupling a transmon qubit to its readout resonator through a Josephson junction with energy $E_{Jc}$, placed in parallel with a capacitive coupling of strength $J$. The circuit Hamiltonian is \cite{Chapple2025}
\begin{align}
    \hat{H} &= \hat{H}_t + \hat{H}_r - E_{Jc} \cos{(\hat{\varphi}_t - \hat{\varphi}_r)} + J \hat{n}_t \hat{n}_r. \label{Eq:Hamiltonian}
\end{align}
Here, $\hat{H}_t = 4 E_C (\hat{n}_t - n_g)^2 - E_J \cos{(\hat{\varphi}_t)}$ is the transmon Hamiltonian with charging energy $E_C$, Josephson energy $E_J$, and gate charge offset $n_g$. The resonator Hamiltonian is $\hat{H}_r = \omega_r \hat{a}^\dagger \hat{a}$ with frequency $\omega_r$. The operators $\hat{n}_t$ ($\hat{n}_r$) and $\hat{\varphi}_t$ ($\hat{\varphi}_r$) denote the charge and phase operators of the transmon (resonator). 
To better understand the benefits of this approach, we expand the interaction terms in \cref{Eq:Hamiltonian} as
\begin{align}
    \hat{H}_{\text{int}} = - E_{Jc} \left[ \cos{\hat{\varphi}_t} \cos{\hat{\varphi}_r} + \sin{\hat{\varphi}_t} \sin{\hat{\varphi}_r} \right] + J \hat{n}_t \hat{n}_r.
    \label{eq:int_hamiltonian}
\end{align}

Writing the phase operators as $\hat{\varphi}_t= \varphi_{\mathrm{zpf,t}}(\hat{b}+\hat{b}^\dagger)$ and  $\hat{\varphi_r}= \varphi_{\mathrm{zpf,r}}(\hat{a}+\hat{a}^\dagger)$, where $\hat{b}$ and $\hat{a}$ are the qubit and resonator annihilation operators, respectively, and $\varphi_{\mathrm{zpf},t}$ and $\varphi_{\mathrm{zpf},r}$ are their corresponding zero-point phase fluctuations, we can expand the interaction to second order in phase fluctuations. To that order, the $\cos{\hat{\varphi}_t}\cos{\hat{\varphi}_r}$ term generates a cross-Kerr interaction of the form $\hat{b}^\dagger \hat{b}\hat{a}^\dagger \hat{a}$, whose amplitude depends only on the coupling junction energy and the zero-point phase fluctuations. Consequently, the amplitude of this interaction is independent of the resonator frequency (hence nonperturbative), in stark contrast to the cross-Kerr interaction that arises from a transverse coupling via a Schrieffer–Wolff transformation \cite{Blais2021}. 

The same $\cos{\hat{\varphi}_t}\cos{\hat{\varphi}_r}$ term also produces a nonperturbative self-Kerr nonlinearity on the resonator proportional to $\hat{a}^\dagger \hat{a}^\dagger\hat{a}  \hat{a}$ and whose strength depends on the resonator impedance \cite{Chapple2025}. As we show later, this induced nonlinearity can be intentionally exploited to perform bifurcation-based readout. 

On the other hand, the $\sin{\hat{\varphi}_t} \sin{\hat{\varphi}_r}$ term produces, to leading order, an undesired transverse coupling proportional to $\hat{\varphi}_t \hat{\varphi}_r \propto \hat{a}^\dagger \hat{b}+\hat{a}\hat{b}^\dagger$. A key feature of junction readout is that this transverse coupling can be mitigated—or even canceled—by choosing an appropriate capacitive coupling strength $J$ such that the exchange matrix element between qubit and resonator vanishes. In other words, one can choose $J$ such that $\langle 1_t, 0_r \vert \hat{H}_{\rm{int}} \vert 0_t, 1_r \rangle \propto \langle 1_t, 0_r \vert \hat{a}^\dagger \hat{b}+\hat{a}\hat{b}^\dagger \vert 0_t, 1_r \rangle = 0$. We refer to this operating point as the balanced condition \cite{Chapple2025}.

Here, we implement junction readout with the circuit schematic shown in \cref{Fig:device Schematic}(a). The transmon's effective Josephson energy $E_J(\Phi_{\rm{ext}})$ is made flux-tunable by an asymmetric superconducting interference device (SQUID). It is coupled to a $\lambda/4$ coplanar-waveguide resonator with frequency $\omega_r$, which is itself coupled to a feedline with decay rate $\kappa$. Moreover, an input capacitor $C_{\rm in}$ is incorporated into the feedline to provide directionality to the readout signal by preferentially routing emitted photons toward the output port \cite{Swiadek2024,Heinsoo_rapid_fidelity}.

As shown in \cref{Fig:device Schematic}(a), this realization of junction readout features three loops. The two large loops, outlined by the red dashed and blue dash-dotted lines, are set by the qubit–resonator geometry, while the small green loop corresponds to the transmon’s SQUID loop. A DC coil beneath the sample provides the flux bias. Flux threading the large loops modulates both $E_J$ and $E_{Jc}$, which in turn modulates the resonator frequency (see \cref{appendix:sample description}). This modulation can be suppressed by biasing at integer flux quanta, enabling selective tuning of only the SQUID loop.

The chip contains four such qubit–resonator pairs, coupled at different positions along the feedline, with varying Josephson energies ($E_{J_c}$, $E_J$) but identical layout (see Appendix \ref{appendix:sample description}). Unless otherwise specified, we focus on one representative pair. Figures~\ref{Fig:device Schematic}(b–d) show an optical micrograph of the pair and Scanning electron microscope images of the transmon, coupling junction, and SQUID. More fabrication and setup details are provided in Appendix~\ref{appendix:device fab and setup}.

\begin{figure}
    \centering
    \includegraphics[width= 1\columnwidth]{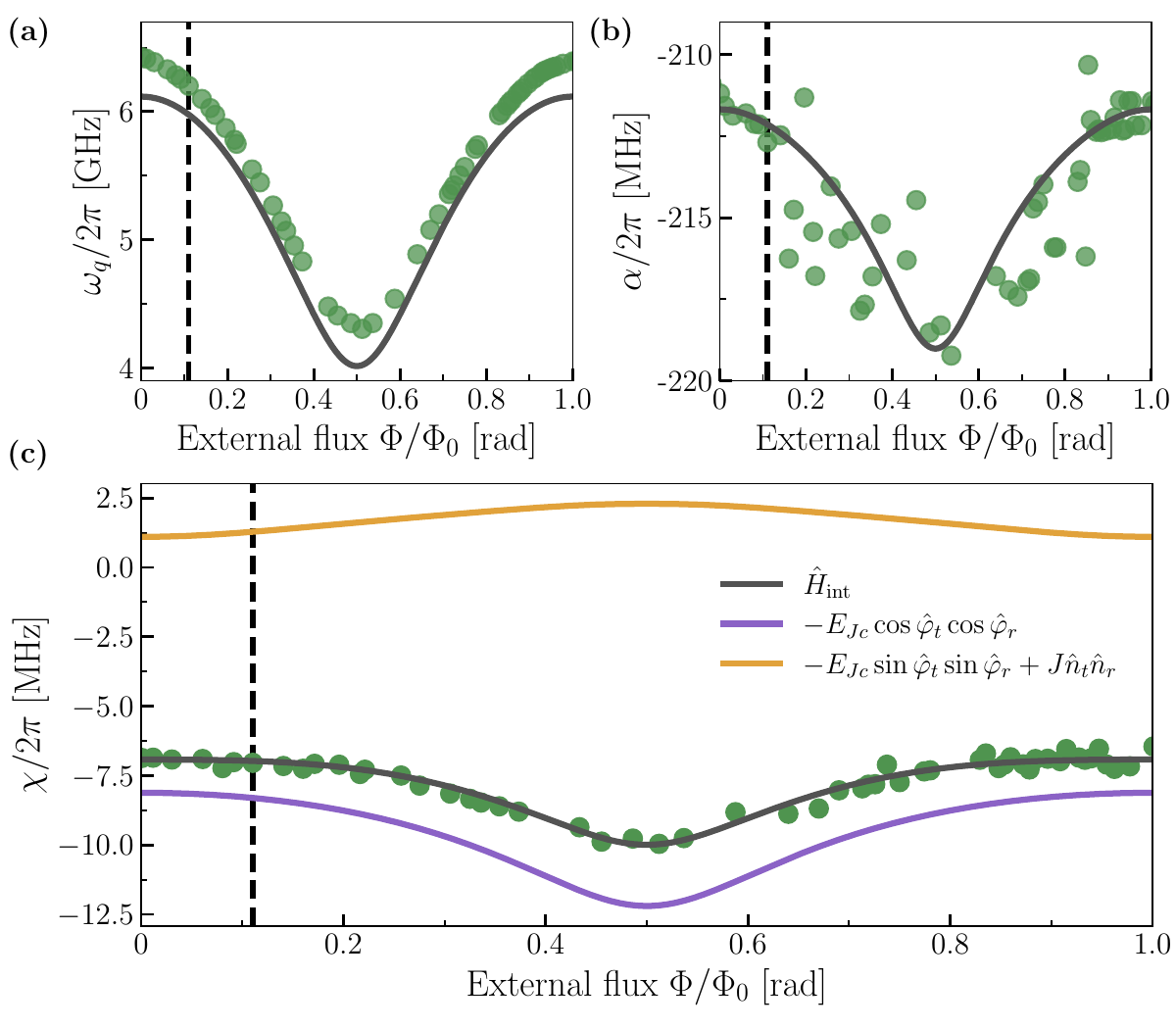}
    \caption{\textbf{Circuit parameters.} (a) Qubit frequency, (b) qubit anharmonicity and (c) cross-Kerr shift as a function of flux threading the transmon's asymmetric SQUID. The solid black lines are fits obtained from \cref{Eq:Hamiltonian} from which we extract $E_{J_c}/2\pi = 4.01$ GHz, $J/2\pi = 83.0$ MHz, and resonator impedance $Z_r = 68.8 \: \Omega$. (d) Calculated contributions of the transverse and nonperturbative coupling terms. In all panels, the vertical dashed line marks the flux bias point used for the readout characterization in the following sections. 
    }
    \label{Fig:flux_response}
\end{figure}

\Cref{Fig:flux_response} shows the qubit frequency $\omega_q$, anharmonicity $\alpha$, and cross-Kerr shift $\chi$ as a function of the external flux threading the transmon SQUID. The cross-Kerr shift is defined as half the difference between the resonator frequencies obtained when the qubit is prepared in $\ket{g}$ and $\ket{e}$. We measure a resonator decay rate of $\kappa/2\pi = 10.6$ MHz (see Appendix~\ref{Appendix:Calibration}), and the resonator frequency $\omega_r / 2\pi$ varies between $7.65$ and $7.66$ GHz over the flux range. 

Solid black lines in \cref{Fig:flux_response}(a) and (b) correspond to numerical fits to \cref{Eq:Hamiltonian} from which we extract $E_{Jc}/2\pi = 4.01$ GHz, $J/2\pi = 83$ MHz, and a resonator impedance of $Z_r = 68.8 \: \Omega$. Furthermore, we find that the SQUID’s total Josephson energy is $E_{{J_ \Sigma}}/2\pi = 21.56$ GHz with an asymmetry of $0.346$ (see \cref{appendix:lumped_model_junction_readout}). We note that the fit is obtained using only the experimentally measured qubit frequencies and anharmonicities. All other quantities are subsequently derived from the fitted parameters with the Hamiltonian in \cref{Eq:Hamiltonian}. 

\Cref{Fig:flux_response}(c) further decomposes the cross-Kerr shift into contributions from the nonperturbative $\cos{\hat{\varphi}_t}\cos{\hat{\varphi}_r}$ interaction (purple line) and from the transverse $\sin{\hat{\varphi}_t}\sin{\hat{\varphi}_r}$ and $\hat{n}_t\hat{n}_r$ couplings (orange line). Because tuning the flux through the SQUID modifies the transmon’s phase zero-point fluctuations, the relative strengths of these contributions vary with flux. As expected, the dominant contribution to the cross-Kerr shift arises from the nonperturbative $\cos{\hat{\varphi}_t}\cos{\hat{\varphi}_r}$ interaction, which is at least five times larger than the combined contribution of the transverse $\sin{\hat{\varphi}_t}\sin{\hat{\varphi}_r}$ and $\hat{n}_t\hat{n}_r$ couplings.

Two points are worth emphasizing. First, the contribution from the combined transverse couplings is positive, which reduces the overall cross-Kerr strength; see \cref{Appendix_cross_kerr_contributions} for details. Second, because the exchange coupling in junction readout is strongly suppressed by the interference between $\sin{\hat{\varphi}_t} \sin{\hat{\varphi}_t} \sim \hat{\varphi}_t \hat{\varphi}_r$ and $\hat{n}_t \hat{n}_r$ interactions, the transverse contribution to the cross-Kerr shift arises predominantly from the remaining two-mode squeezing term proportional to $\hat{a}\hat{b} + \hat{a}^\dagger\hat{b}^\dagger$. This is further confirmed in \cref{Appendix_cross_kerr_contributions}, where we decompose the transverse part of the cross-Kerr using a Schrieffer–Wolff perturbation expansion into contributions from the exchange (co-rotating) and the two-mode-squeezing (counter-rotating) matrix elements, proportional to $\hat{a}^\dagger \hat{b} + \hat{a}\hat{b}^\dagger$ and $ \hat{a}\hat{b} + \hat{a}^\dagger\hat{b}^\dagger$, respectively. It is worth noting that in standard dispersive readout, suppressing the exchange coupling would largely eliminate the dispersive shift. In contrast, junction readout retains a large cross-Kerr shift even when the exchange term is canceled.

Finally, as shown in \cref{appendix:charactherization of the device parameter}, measurements across the different resonators on the chip confirm that the cross-Kerr shift scales with the coupling-junction energy $E_{Jc}$.

\section{Intrinsic Purcell Protection and critical photon numbers} 
\label{sec:purcell_protection_ncrits}

Another distinctive feature of junction readout is its intrinsic Purcell filtering, which arises from the coupling made of a capacitor in parallel with a Josephson junction naturally forming a lumped-element LC notch filter. Indeed, when the qubit frequency coincides with the notch frequency $\omega_n = 1/\sqrt{L_{J_c} C_c}$, where $L_{J_c}=\left(\hbar/2e\right)^{2}/E_{J_c}$ is the linearized inductance of the coupler Josephson junction and $C_c$ is the coupling capacitance,  the filter presents a high impedance, thereby suppressing energy relaxation of the qubit through the resonator and, equivalently, maximizing its Purcell-limited lifetime $T_1^{\rm pl}$.
 
Since the qubit relaxation time $T_1$ contains contributions from multiple decay channels beyond Purcell loss, measuring $T_1$ as a function of qubit frequency alone does not provide a direct characterization of the Purcell filtering. Instead, we follow the method of Ref.~\cite{Sunada2022}, which provides an accurate estimate of the Purcell-limit of the lifetime, $T_1^{\rm pl}$, even when the qubit is not Purcell limited. In this approach, the qubit is driven through the resonator feedline with power $P$ at its transition frequency, inducing resonant Rabi oscillations with frequency $\Omega$. The Purcell-limited lifetime is then given by \cite{Sunada2022}
\begin{equation}
\label{Eq:external_coupling}
T_{1}^{\rm pl} = \frac{4P}{\Omega^{2}\hbar \omega_q}.
\end{equation}

To calibrate the drive power at the device input, we follow the approach of Ref.~\cite{Yen2025}. When the qubit is biased at the flux point where its frequency is maximally detuned from the notch filter frequency, the relaxation is expected to be dominated by Purcell decay, so that $T_1^{\rm pl} \approx T_1$. Thus, the applied drive power can be estimated as $P \approx \Omega^{2}\hbar \omega_q T_1/4$.

\begin{figure}
    \centering
    \includegraphics[width= 1\columnwidth]{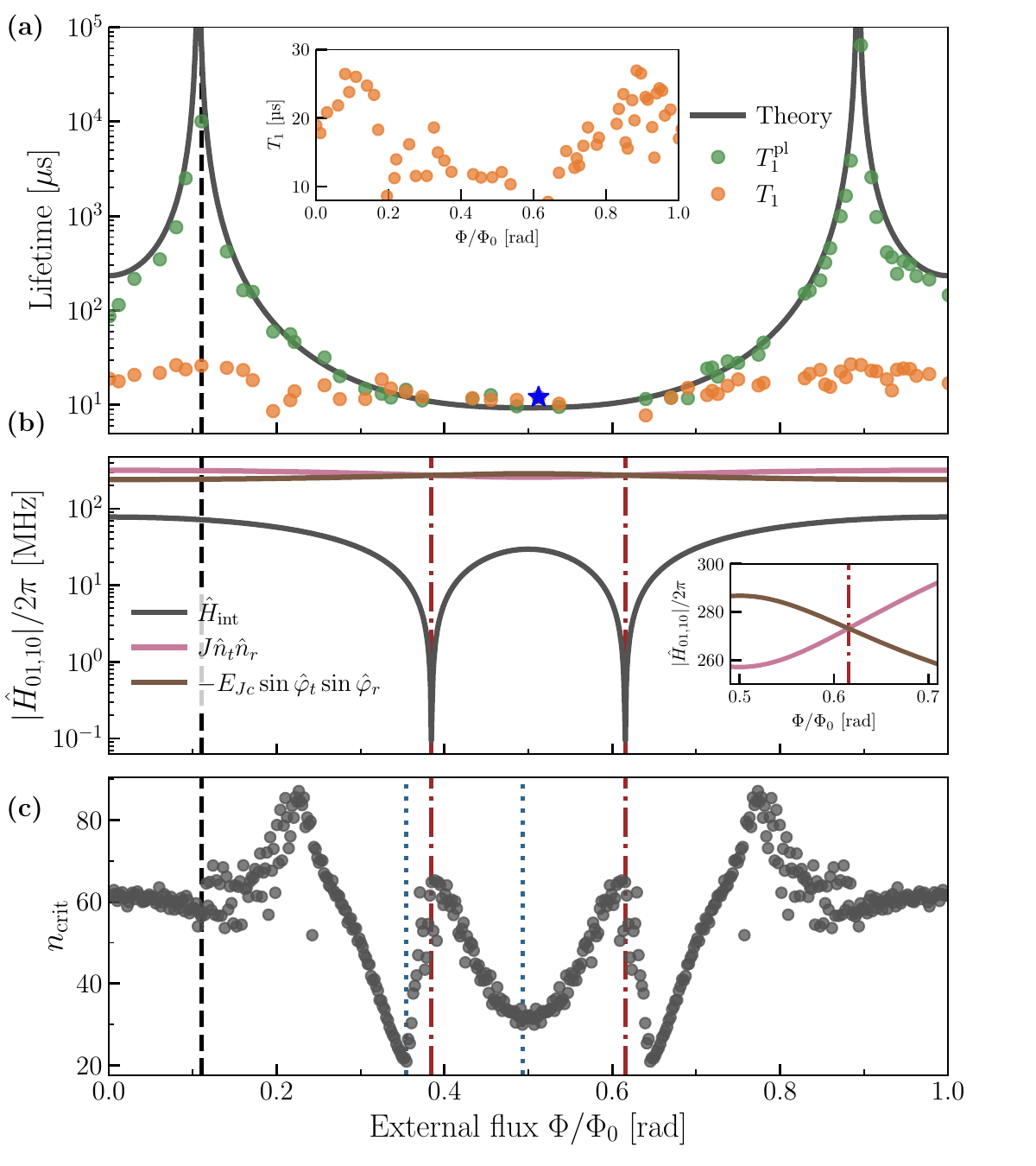}
    \caption{\textbf{Intrinsic Purcell protection and critical photon numbers.} (a)  Purcell-limited relaxation time $T_1^{\rm pl}$ (green circles) and relaxation time $T_1$ (orange circles) as a function of flux threading the transmon SQUID. The inset shows a zoom of the $T_1$ revealing two distinct peaks corresponding to the notch frequency encountered twice over one flux quantum. The solid black line shows the theoretical prediction obtained from master-equation simulations using the fitted parameters from \cref{Fig:flux_response}. The blue star marks the operating point used for power calibration, while the gray vertical dashed line indicates the flux bias at which Purcell filtering is maximal and which is used for the readout characterization in subsequent sections.
    (b) Absolute value of the exchange matrix element $\vert \bra{0,1}\hat{H}_{\mathrm{int}} \ket{1,0} \vert$, illustrating the suppression of the qubit–resonator exchange coupling due to cancellation between the
    the $\hat{n}_t \hat{n}_r$ and $\sin{\hat{\varphi}}_t \sin{\hat{\varphi}}_r$ interaction terms. The red vertical dashed–dotted line indicates the balanced-coupling condition. 
    (c) Onset of measurement-induced state transitions $n_{\mathrm{crit}}$
    as a function of external flux. Blue dotted vertical lines indicate multiphoton resonances, which give rise to sharp drops in $n_{\mathrm{crit}}$. The gray vertical dashed line again marks the flux point of maximal Purcell protection, while the red dashed–dotted lines denote the balanced-coupling point.
    }
    \label{Fig:intrisic_purcell}
\end{figure}

\Cref{Fig:intrisic_purcell}(a) shows the Purcell-limited relaxation time $T_1^{\rm pl}$ (green circles) and the measured relaxation time $T_1$ (orange circles) as a function of the flux threading the transmon SQUID. As the qubit frequency approaches the notch frequency (black vertical dashed line), $T_1^{\rm pl}$ increases by nearly four orders of magnitude. Two peaks appear as the flux bias spans one flux quantum, bringing the qubit frequency through the notch twice. The solid black line shows the theoretical Purcell decay rate obtained from master equation simulations using the parameters extracted from the fits in~\cref{Fig:flux_response}; see~\cref{appendix:purcell_theory} for more details. We observe excellent agreement between theory and experiment. 

The inset highlights the measured $T_1$ lifetimes, which follows the same trend as $T_1^{\rm pl}$ away from the notch but saturates near $T_1 = 27~\mu{\rm s}$ when the qubit frequency is close to the notch frequency, reflecting the presence of other relaxation channels. The blue star indicates the point used for calibrating the drive power, where the qubit is indeed Purcell limited. In~\cref{appendix:purcell protection}, we show Purcell lifetimes obtained from finite-element electromagnetic simulations, which also exhibit excellent agreement with the measured lifetimes. Finally, at the point of maximum Purcell filtering (black vertical dashed line), we measure a Hahn-echo pure dephasing time of $T_2^{E}=11~\mu{\rm s}$ and an inhomogeneous dephasing time of $T_2^{*}=4~\mu{\rm s}$, extracted from Ramsey measurements.

It is important to note that the flux point where $T_1^{\rm pl}$ is maximized does not coincide with the balanced-coupling point, where the qubit–resonator exchange interaction is suppressed. To illustrate this explicitly, \cref{Fig:intrisic_purcell}(b) shows the $\ket{0_t,1_r} \leftrightarrow \ket{1_t,0_r}$ matrix element of the interaction Hamiltonian $\hat{H}_{\mathrm{int}}$, together with the separate contributions coming from the $\hat{n}_t \hat{n}_r$ and $\sin{\hat{\varphi}}_t \sin{\hat{\varphi}}_r$ couplings. One can see that these two contributions cancel at $\Phi \approx 0.384\Phi_0$ (red dashed-dotted line), where $\vert \hat{H}_{01,10} \vert = \vert \bra{0_t,1_r}\hat{H}_{\mathrm{int}}\ket{1_t,0_r}\vert $ approaches zero (see inset). In contrast, the Purcell filtering is maximal at around $\Phi \approx 0.107\Phi_0$ (black dashed line).

The origin of this distinction lies in the fact that, even when the exchange coupling is canceled at the balanced condition, Purcell decay can still occur through the two-mode-squeezing interaction. As shown in \cref{appendix:purcell_theory}, in the dispersive limit the Purcell decay rate can be approximated as
\begin{equation}
\label{main: T_1 equation}
\frac{1}{T_1^{\rm pl}}
= \kappa \left(
\frac{\bra{0_t,1_r} \hat{H}_{\mathrm{int}}\ket{1_t,0_r}}{\omega_r - \omega_q}
+
\frac{\bra{0_t,0_r} \hat{H}_{\mathrm{int}}\ket{1_t,1_r}}{\omega_r + \omega_q}
\right)^{2}.
\end{equation}
The first term of this expression arises from the exchange interaction and vanishes at the balanced-coupling point, whereas the second term originates from the two-mode-squeezing interaction and does not vanish at that point. In standard dispersive readout architectures that rely only on the transverse $\hat{n}_t \hat{n}_r$ coupling, the magnitudes of these two matrix elements are identical (and opposite in sign). As $\omega_r + \omega_q > |\omega_r - \omega_q|$, their contributions cannot cancel, and therefore no Purcell-cancellation point exists.

In contrast, in the junction readout scheme, the exchange coupling is intentionally engineered to be smaller than the two-mode-squeezing term. This enables destructive interference between the two contributions in Eq.~\eqref{main: T_1 equation}, making Purcell cancellation possible. Importantly, this cancellation occurs precisely when the qubit frequency matches the notch frequency of the LC filter, see \cref{appendix:purcell_theory}. 

As the coupling contributions vary with external flux, the onset of measurement-induced state transitions—which we refer to as the critical photon number $n_{\mathrm{crit}}$—is likewise expected to display a rich flux dependence. This behavior is indeed observed in \cref{Fig:intrisic_purcell}(c). Here, $n_{\mathrm{crit}}$ is computed using the branch-analysis method of Ref.~\cite{Shillito2022}, a numerical procedure for labeling the dressed states of the qubit–resonator system that enables identification of the resonator photon number above which multiphoton resonances involving the computational states appear in the spectrum \cite{Shillito2022, Dumas2024}. These resonances typically involve higher excited states of the transmon near or above the cosine potential which are sensitive to gate charge $n_g$. 
Since gate charge can fluctuate from shot to shot in experiments, the critical photon numbers shown in \cref{Fig:intrisic_purcell}(c) are averaged over gate charge in the range $n_g \in [0, 0.5]$. 

The drops in $n_{\mathrm{crit}}$, indicated by the vertical blue dotted lines, arise from multiphoton resonances involving the absorption of one or more photons in the readout resonator. Notably, the balanced-coupling point (red dashed line) features a relatively large critical photon number, especially considering the large cross-Kerr strength at this flux bias ($2\chi/2\pi \sim -24~\mathrm{MHz}$), in agreement with Ref.~\cite{Chapple2025}. 

Interestingly, the flux bias point at which the Purcell filtering is maximized (grey dashed line) also exhibits a large critical photon number, $n_{\mathrm{crit}} \sim 60$. Combined with the strong cross-Kerr interaction at this flux bias point, $2\chi/2\pi \sim -14$ MHz, this offers an excellent operating point for readout with minimal leakage. In the following, we fix the flux bias to this value ($\Phi \approx 0.107 \Phi_0$) to experimentally explore the readout landscape in detail. With appropriate optimization, we further demonstrate that the junction readout architecture supports fast, high-fidelity, and highly QND measurement.

\section{Readout Performance} 

The resonator Kerr nonlinearity introduced by the coupling junction enables bifurcation dynamics that we exploit for qubit readout. When the drive amplitude and detuning exceed a critical threshold, the nonlinear resonator switches from having a single stable steady state to exhibiting two distinct stable states corresponding to low- and high-photon-number amplitudes \cite{Siddiqi2004,Siddiqi2005,Mallet2009}. In our implementation, the large cross-Kerr interaction produces a strong qubit-state-dependent separation of the resonator response, while the partial cancellation of the transverse interaction allows operation at high photon number without detrimental effects such as MIST. As a result, the qubit state can be mapped onto the two dynamical steady states of the nonlinear resonator, enabling high-fidelity bifurcation-based readout.

\begin{figure}
    \centering
    \includegraphics[width= \columnwidth]{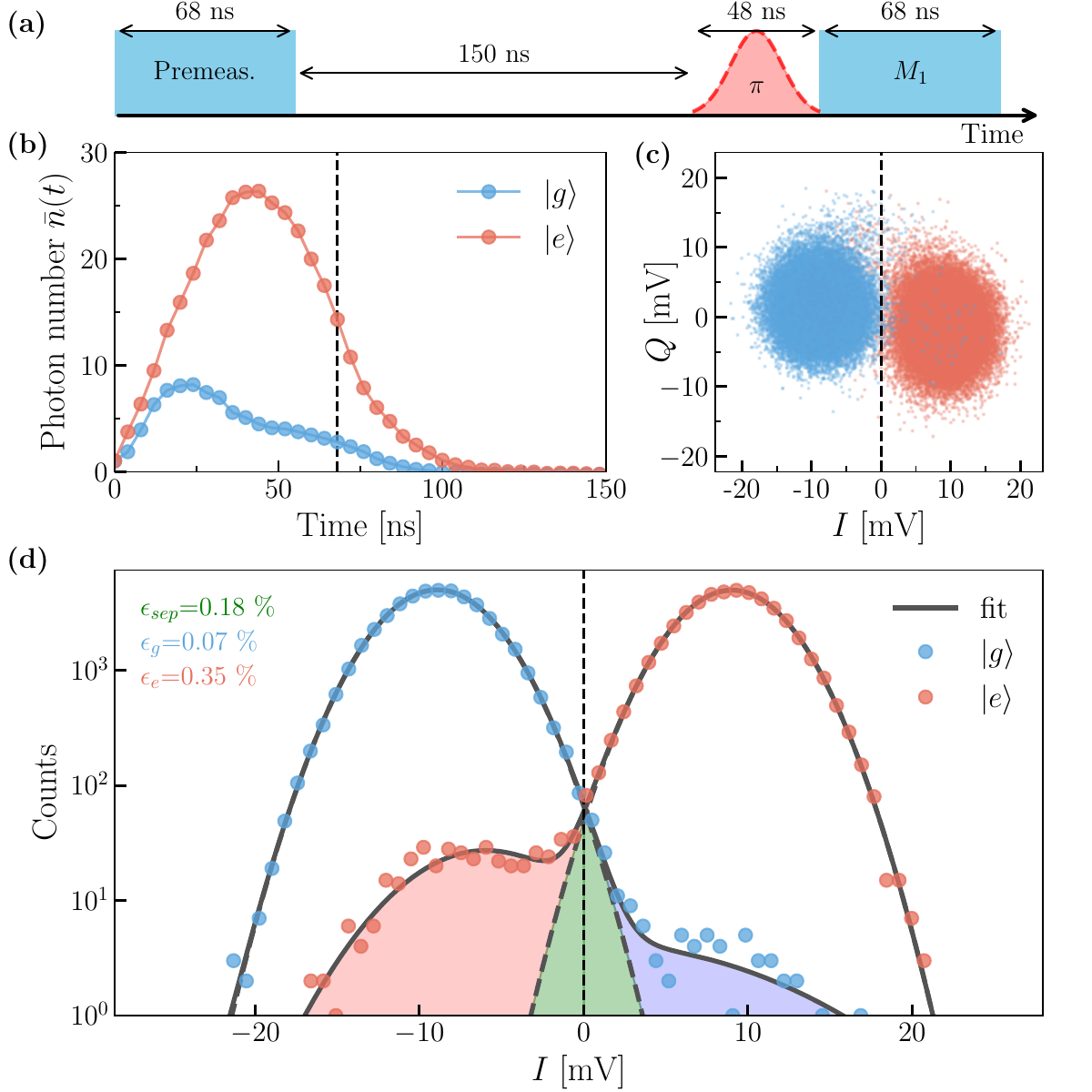}
    \caption{\textbf{Single-shot readout performance}. (a)  Pulse sequence used for readout fidelity characterization: a preselection pulse heralds the qubit to the ground state, after which the qubit is optionally excited and read out. (b) Time evolution of the intracavity photon number for the qubit in $\ket{g}$ (blue) and $\ket{e}$ (red). The dashed line marks the end of the readout pulse and integration window. (c) $5 \times 10^{4}$ single-shot readout events for when the qubit is prepared in the excited (red) and ground (blue) states. The dashed line indicates the linear discriminant used to calculate the assignment fidelity. (d) Histograms of the measured quadrature distributions in (c). The solid black lines show fits to a double Gaussian distribution. The blue and red shaded regions indicate readout errors arising from qubit decay or excitation during the measurement. Taking half of each shaded area gives their contribution to the total error ($\epsilon_e = 0.35$\,\% and $\epsilon_g = 0.07$\,\%). The overlap of the two distributions (half of the green shaded area), contributes a state-separation error of $\epsilon_{\rm sep} = 0.18$\,\%. The resulting assignment fidelity is $99.4$\,\% with an integration time of $68$\,ns.}
    \label{Fig:ReadFidelity}
\end{figure}

In this section, we focus on the readout performance acheived when  the drive amplitude and frequency are optimized for both fidelity and QNDness.
The following section examines how these metrics vary with drive frequency and amplitude while keeping the flux bias fixed. The readout performance is characterized using the protocol sketched in Fig.~\ref{Fig:ReadFidelity}(a). The sequence begins with a $68$-ns preselection pulse that verifies that the qubit is in the ground state \cite{Johnson2012,Walter2017}. After a waiting time of $\tau_{\rm w}=150$\,ns—sufficient to passively reset the resonator ($\kappa \tau_{\rm w} \approx 10$)—the qubit is either left in $\ket{g}$ or excited to $\ket{e}$ using a $48$-ns  Gaussian derivative-removal-by-adiabatic-gate (DRAG) pulse \cite{Motzoi2009,Gambetta2011}. A $68$-ns square readout pulse is then applied to the resonator, and the resulting measurement signal is integrated with optimal weights \cite{Walter2017,Bultink2018} to maximize discrimination between the two qubit states.

The quantum efficiency of the readout chain, $\eta \approx 4$\,\%, is measured in the resonator's linear regime (see Appendix~\ref{Appendix_Quantum_efficiency}) by comparing the measurement-induced dephasing to the measured SNR \cite{Bultink2018}. The low efficiency results from using a 4K HEMT as a first-stage amplifier, which sets a relatively high noise floor for the readout chain. Nevertheless, as shown below, we achieve high readout fidelity due to the large cross-Kerr interaction and the comparatively high intracavity photon population, which remains well below the critical photon number. This enables a strong, state-dependent resonator response despite the modest quantum efficiency.

Figure \ref{Fig:ReadFidelity}(b) shows the intracavity photon number, extracted from the measured ac-Stark shift (see Appendix \ref{Appendix:Calibration}), for the qubit prepared in either $\ket{g}$ or $\ket{e}$. Owing to the resonator's nonlinearity, the dynamics differ substantially between the two qubit states. High fidelity readout is achieved in the region where $\ket{g}$ is encoded in the low-photon branch and $\ket{e}$ in the high-photon branch. We observe that the intracavity photon number reaches a maximum of $26$ and $8$ photons when the qubit is initialized in $\ket{e}$ and $\ket{g}$, respectively. The black dashed line marks the end of the $68$-ns readout pulse and integration window. We note that the intracavity photon number begins to decrease before the end of the pulse. This occurs because the pulse induces transient oscillations in the intracavity field which causes the photon number to overshoot and ring before settling to its steady-state amplitude.

The results of $5 \times 10^{4}$ readout events are shown in Fig.~\ref{Fig:ReadFidelity}(c), with blue (red) dots corresponding to the transmon prepared in $\ket{g}$ ($\ket{e}$). A linear discriminant (black dashed line) is used to separate the two clusters. The assignment fidelity is computed as $\mathcal{F}\equiv[P_0(g|g)+P_{\pi}(e|g)]/2$, where $P(a|b)$ denotes the conditional probability of measuring the outcome $a$ given the preselection outcome $b$, and the labels $\pi$ and $0$ indicate sequences performed with and without a $\pi$ pulse on the qubit. We obtain an assignment fidelity of $99.4$\,\% with an integration time of $68$ ns. 

Projecting the measurements onto the $I$ quadrature gives the histograms in Fig.~\ref{Fig:ReadFidelity}(d). Both histograms are well described by a double Gaussian distribution fit (solid lines) \cite{Johnson2012, Walter2017}. From the overlap of the two fitted distributions (half of the green shaded area), we extract a state-separation error of $\epsilon_{\rm sep}=0.18$\,\%. Additional errors arise from qubit-state transitions during the readout, contributing to $\epsilon_g=0.07$\,\% for $\ket{g}$ (half of the blue shaded area) and $\epsilon_e=0.35$\,\% for $\ket{e}$ (half of the red shaded area). We estimate the relaxation-induced contribution to $\epsilon_e$ during the measurement to be $\epsilon_{\rm cl} =\tau_m /(2T_1) \approx 0.13$\,\%, using $T_1 = 27 \: \mu$s and the measurement integration time $\tau_m = 68$ ns. Incorporating a near-quantum-limited amplifier at the first stage is expected to primarily reduce the state-separation error $\epsilon_{\rm sep}$. The remaining error is attributed to imperfections in the preselection measurement and excited-state preparation, as well as switching events between the two dynamical resonator states during measurement—a known effect in bifurcation-based readout \cite{Mallet2009,Vijay2009,Andersen2020} (see \cref{Appendix_QNDness_error} for details). A more comprehensive study of these limitations is left for future work.

\section{Parameter Dependence of Readout Fidelity and QNDness}
\label{sec:fidelity_and_QNDness}

\begin{figure}[t]
    \centering
    \includegraphics[width= \columnwidth]{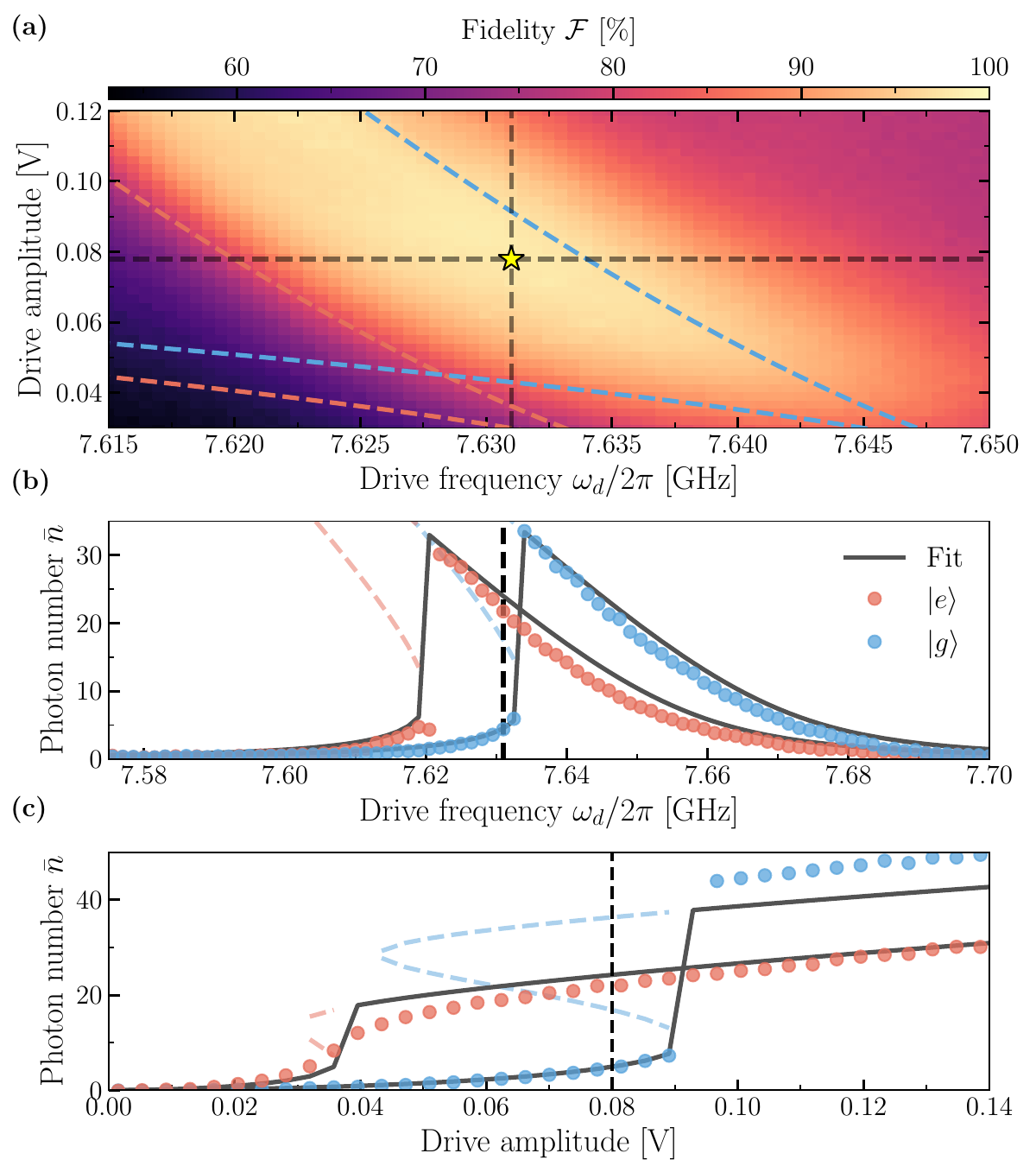}
    \caption{\textbf{Readout fidelity landscape}. (a) Heat map of the readout fidelity as a function of drive frequency $\omega_d$ and readout drive amplitude. The star indicates the parameter used for  Fig.~\ref{Fig:ReadFidelity}. The dashed colored line indicate the resonator bistable region for when the qubit is in state $\ket{g}$ (blue) and $\ket{e}$ (red). The gray dashed lines correspond to the drive frequency and amplitude used in the subsequent panels. (b,c) Intracavity photon number as a function of drive frequency (b) and drive amplitude (c), extracted from the Stark-shift calibration. Solid lines show fits to the stable solution of Eq.~\ref{eq:nonlinear dynamics}, while light dashed lines indicate the regions of unstable solutions. Vertical black dashed lines indicate (b) the drive frequency and (c) the drive amplitude at the star marked parameters in (a).
    }
    \label{Fig:ReadFidelityOptimal}
\end{figure}

We now turn to a discussion to the effect of variations in the drive frequency and amplitude on the readout fidelity, the QND character of the measurement, and the resonator responses.

\Cref{Fig:ReadFidelityOptimal}(a) shows the readout fidelity as a function of the drive frequency $\omega_d$ and drive amplitude, obtained using the protocol described in the previous section. The red and blue dashed lines mark the bistable regions of the resonator for the qubit prepared in $\ket{e}$ and $\ket{g}$, respectively. We find a broad operating window in which the fidelity exceeds $95$\,\%. The operating point used in~\cref{Fig:ReadFidelity} is indicated by the star marker.

\Cref{Fig:ReadFidelityOptimal}(b) and (c) display the steady-state intracavity photon number—determined via ac Stark-shift calibration \cite{Ong2011}—as a function of drive frequency and amplitude for the qubit in $\ket{g}$ (blue) and $\ket{e}$ (red). The calibration procedure is described in \cref{Appendix:Calibration}. 
The solid black lines correspond to fits to the semiclassical steady-state response of a driven nonlinear resonator, described by
\begin{equation}
\label{eq:nonlinear dynamics}
i(\omega_{e/g}-\omega_d)\alpha
+ iK |\alpha|^2 \alpha
+ \frac{\kappa}{2}\alpha
= F,
\end{equation}
where $\omega_{e/g}=\omega_r \pm \chi$ is the qubit-state–dependent resonator frequency (see \cref{appendix:charactherization of the device parameter}), $\omega_d$ is the drive frequency, and $F$ is the drive amplitude. Here, $F$ is the only free parameter, while the self-Kerr nonlinearity is fixed to the state-averaged value $K/2\pi = -0.963$ MHz, obtained from the Hamiltonian \cref{Eq:Hamiltonian} using the previously extracted parameters. The remaining parameters $\kappa$, $\omega_r$, and $\chi$ are set to their experimentally determined values. The dashed vertical lines indicate the drive frequency and amplitude used for the readout-fidelity measurements.

\begin{figure}
    \centering
    \includegraphics[width= \columnwidth]{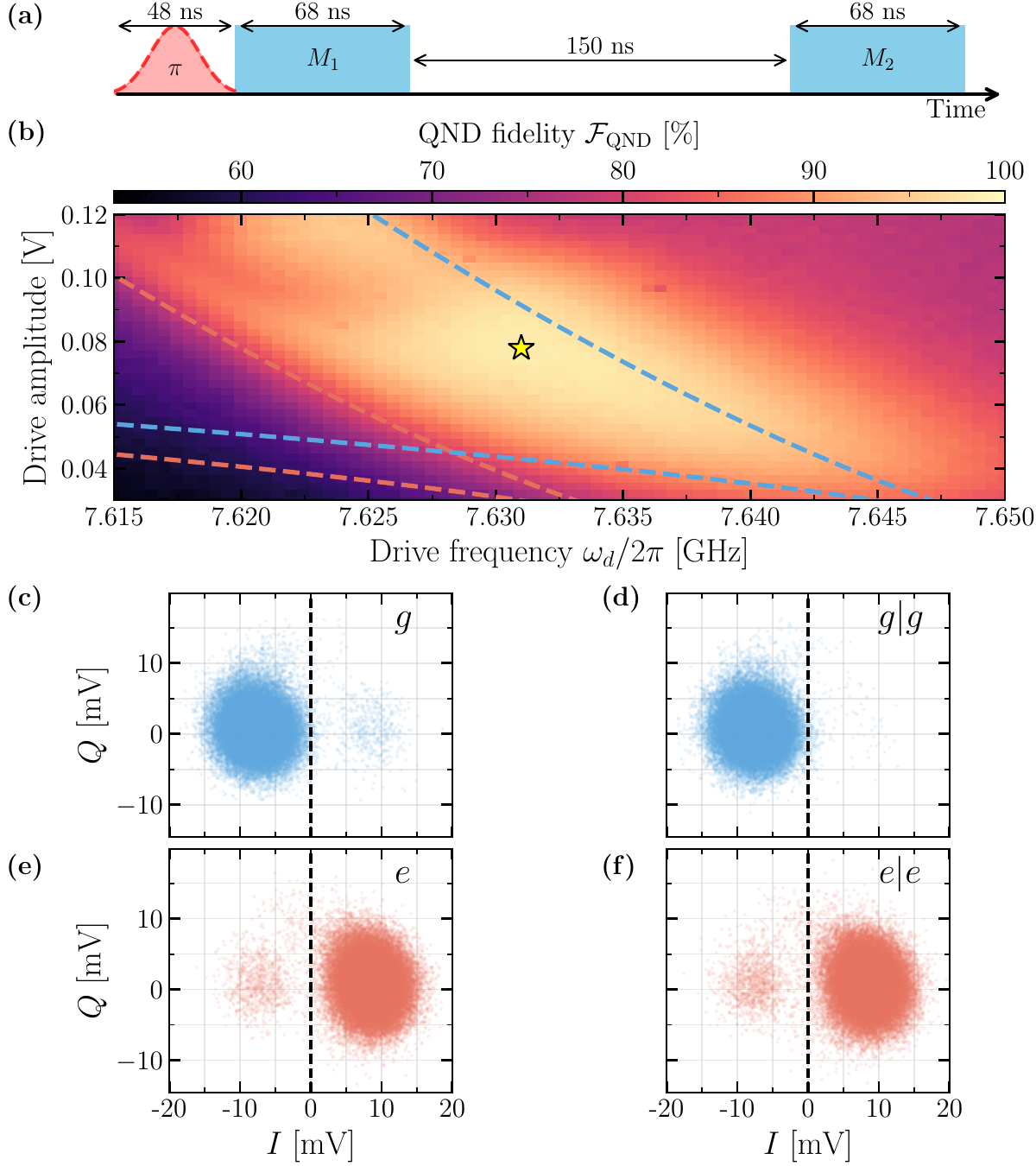}
    \caption{\textbf{QND fidelity landscape}. (a) Pulse sequence used to measure the QND fidelity. The qubit is either left in the ground state $\ket{g}$ or excited to $\ket{e}$ with a $48$-ns $\pi$ pulse, followed by two identical, consecutive $68$-ns-long square readout pulses, separated by $150$\, ns. (b) Heat map of the QND fidelity as a function of drive frequency $\omega_d$ and amplitude. The star marker indicates the operating point used to characterize the readout fidelity, see \cref{Fig:ReadFidelityOptimal}(a). (c) IQ distribution of $5 \times 10^4$ measurement for the first readout with the qubit prepared in $\ket{g}$. (d) IQ distribution for the second measurement, conditioned on the first outcome being $\ket{g}$ $(g|g)$. (e) Same as (c), but with the qubit prepared in $\ket{e}$. (f) Same as (d), but conditioned on the first outcome being $\ket{e}$ ($e|e$). 
    }
    \label{Fig:QND_main}
\end{figure}

We now turn to assessing the QNDness of the measurement. The sequence used to quantify the QND character consists of two consecutive readout pulses separated by $150$\,ns, with the qubit initially prepared either in the ground or excited state, as illustrated in \cref{Fig:QND_main}(a). Between successive experiments, we wait long enough to ensure that the transmon has relaxed to its ground state. We define the QND fidelity as $\mathcal{F}_{\rm QND}\equiv [P_0(g|g)+P_{\pi}(e|g)]/2$, which quantifies the probability that the qubit state remains unchanged by the measurement. To extract these probabilities, we post-select the data based on the outcome of the first measurement. For example, when estimating $P_{\pi}(e|g)$, we retain only those trials in which the first measurement reports $\ket{e}$ and discard those that return $\ket{g}$. 
 
\Cref{Fig:QND_main}(b) shows the QND fidelity as a function of the drive frequency $\omega_d$ and drive amplitude, exhibiting a trend similar to the assignment fidelity in \cref{Fig:ReadFidelityOptimal}(a). We find a broad region of parameter space in which the QND fidelity exceeds $95$\,\%. At the operating point used for the readout-fidelity measurement (indicated by the star), the QND fidelity reaches $98.4$\,\%.
Comparing the IQ distributions from the first measurement, shown in \cref{Fig:QND_main}(c) and (e), with those from the second measurement, \cref{Fig:QND_main}(d) and (f), reveals no significant leakage. This is further supported in \cref{Appendix_QNDness_error}, where we perform a three-state readout discrimination—distinguishing $\ket{g}$, $\ket{e}$, and $\ket{f}$—and show that leakage into the $\ket{f}$ state has only a minimal impact on the extracted QND fidelity. We estimate the contribution to the QND error from energy relaxation to be $\epsilon_{\rm r}^{\rm{QND}}=\frac{\tau_{w}}{2T_1}+\frac{\tau_m}{T_1} \approx 0.53$\,\%, where $\tau_{w}$ is the waiting time between the two pulses and $\tau_m$ is the measurement integration time. We suspect that the remaining QND error arises from switching events between the two stable resonator states, as detailed in \cref{Appendix_QNDness_error}, where we show in more detail how bistability limits the QND error.

\section{Conclusion}  

In this work, we have experimentally demonstrated junction readout—a simple modification of the standard dispersive circuit in which a Josephson junction is added in parallel with the capacitive coupling between a transmon and its readout resonator \cite{Chapple2025, Wang2025}. This hybrid interaction simultaneously generates a strong, nonperturbative cross-Kerr coupling and forms a lumped-element notch filter that provides intrinsic Purcell protection. Moreover, destructive interference between the $\sin{\hat{\varphi}_t} \sin{\hat{\varphi}_r}$ and $\hat{n}_t \hat{n}_r$ couplings reduces the transverse interaction, suppressing measurement-induced leakage. The coupling junction also introduces sufficient nonlinearity in the resonator to enable bifurcation-based readout at relatively high intracavity photon numbers, with a large cross-Kerr strength of $2\chi/2\pi \sim -14$ MHz. The experimental observations are in excellent agreement with the theoretical model.

Even without external Purcell filters or near-quantum-limited amplifiers—and despite the limited quantum efficiency of our readout chain ($\eta \sim 4\%$)—we achieve an assignment fidelity of $99.4$\,\% with a $68$-ns integration window and a QND fidelity of $98.4$\,\%. In both speed and fidelity, the performance ranks among the best reported for superconducting-qubit readout, while requiring significantly less hardware overhead.

This work establishes junction readout as a practical and hardware-efficient alternative to conventional dispersive readout. Importantly, junction readout eliminates the need for dedicated Purcell filters, remains fully compatible with multiplexing, and—as demonstrated here—enables fast, high-fidelity readout even without using near-quantum-limited parametric amplifiers. Moreover, the present work explores only a subset of the rich parameter space available in junction readout. For example, the same architecture can also support conventional linear readout beyond the bifurcation regime studied here, and the induced resonator nonlinearity can be tuned by adjusting the resonator impedance through the placement of the coupling junction \cite{Chapple2025}. Future implementations may benefit from reducing the size of the qubit–resonator flux loop, or from eliminating it altogether by using a single junction to realize both the qubit and the coupling \cite{Chapple2025}. Finally, it would be interesting to explore junction readout in alternative operating regimes, such as high-frequency readout.

\begin{acknowledgments}

\end{acknowledgments}

The device was fabricated in the Center of MicroNanoTechnology
(CMi) at EPFL. We thank V. Jouanny, L. Peyruchat and F. Minganti for useful discussions. P.S. acknowledges support from the Swiss National Science Foundation (SNSF), through Grant number 200021\_200418 / 1 ,and from the Swiss State Secretariat for Education, Research and Innovation (SERI) under contract number UeM019-16 and MB22.00081 / REF-1131 -52105. M.S.
acknowledges support from the EPFL Center for Quantum Science and Engineering postdoctoral fellowship.
O.\,B.\,K., A.\,A.\,C., and A.\,B.\ were supported by NSERC, the Minist\`ere de l'\'Economie et de l'Innovation du Qu\'ebec, the Canada First Research Excellence Fund, and the U.S. Department of Energy, Office of Science, National Quantum Information Science Research Centers, Quantum Systems Accelerator.\\

\section*{Author Contributions}

G.B., M.S., and P.S. devised the research project. G.B. and J.C. designed the experiment. G.B. fabricated the device. G.B. and J.C. performed the measurements. A.A.C. and O.B.K. numerically simulated the model Hamiltonian. A.A.C. and O.B.K. performed the master-equation simulations used to evaluate the intrinsic Purcell protection, with input from P.A.S.. G.B., J.C., M.S., A.A.C., and O.B.K. analyzed the data. P.S. and A.B. supervised the experimental and theoretical aspects of the project, respectively.

\appendix
\vspace{0.4in}

\newcommand{\IN}[1]{\hat{#1}_{\rm in}}
\newcommand{\OUT}[1]{\hat{#1}_{\rm out}}

\section{Detailed sample description}
\label{appendix:sample description}

\begin{figure}[!h]
    \centering
    \includegraphics[width= 1\columnwidth]{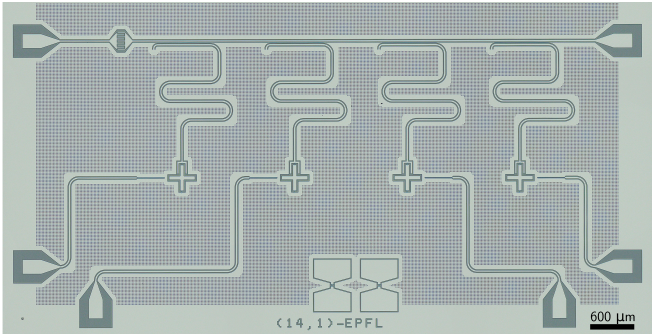}
    \caption{\textbf{Device overview}. Optical micrograph of the device containing four qubit–resonator pairs coupled to a common feedline.}
    \label{sup fig:full chip}
\end{figure}

A micrograph of the full device is shown in \cref{sup fig:full chip}. The chip contains four qubit-resonator pairs that share the same layout with differing values of transmon's intrinsic Josephson energy $E_J$ and the coupling Josephson junction energy $E_{Jc}$. Starting from the left, we label the qubits $\textrm{Q}_1$ through $\textrm{Q}_4$, with each $\textrm{Q}_i$ coupled to its dedicated resonator. From $\textrm{Q}_1$ to $\textrm{Q}_4$, the corresponding qubit–resonator coupling Josephson energies $E_{Jc}$ decrease. In the main text we focus on $\textrm{Q}_2$ which serves as a representative example. In this section we provide additional characterization data for the remaining qubit-resonator pairs. The measured parameters of the four pairs at zero flux are summarized in Table~\ref{table:chip_parameters}.

\begin{table}[h!]
    \centering
    \caption{\textbf{Measured qubit and resonator parameters at zero external flux.}
    Here, $\omega_r$ and $\kappa$ are the resonator frequency and coupling rate to the feedline, respectively,
    $\omega_q$ is the qubit frequency, $\alpha$ is the anharmonicity, and $\chi$ is the cross-Kerr shift.}
    \label{table:chip_parameters}

    \begin{tabular*}{\columnwidth}{@{\extracolsep{\fill}}lccccc@{}}
        \toprule
        &
        \makecell{$\omega_r/2\pi$ \\ (GHz)} &
        \makecell{$\kappa/2\pi$ \\ (MHz)} &
        \makecell{$\omega_q/2\pi$ \\ (GHz)} &
        \makecell{$\alpha/2\pi$ \\ (MHz)} &
        \makecell{$\chi/2\pi$ \\ (MHz)} \\
        \midrule
        $\textrm{Q}_1$ & 7.80 & 4.95 & 6.62 & $-209$ & $-7.62$ \\
        $\textrm{Q}_2$ & 7.66 & 10.6 & 6.45 & $-210$ & $-6.83$ \\
        $\textrm{Q}_3$ & 7.55 & 3.98 & 6.11 & $-213$ & $-6.17$ \\
        $\textrm{Q}_4$ & 7.44 & 3.68 & 5.74 & $-215$ & $-6.06$ \\
        \bottomrule
    \end{tabular*}
\end{table}

\begin{figure}[!h]
    \centering
    \includegraphics[width= 1\columnwidth]{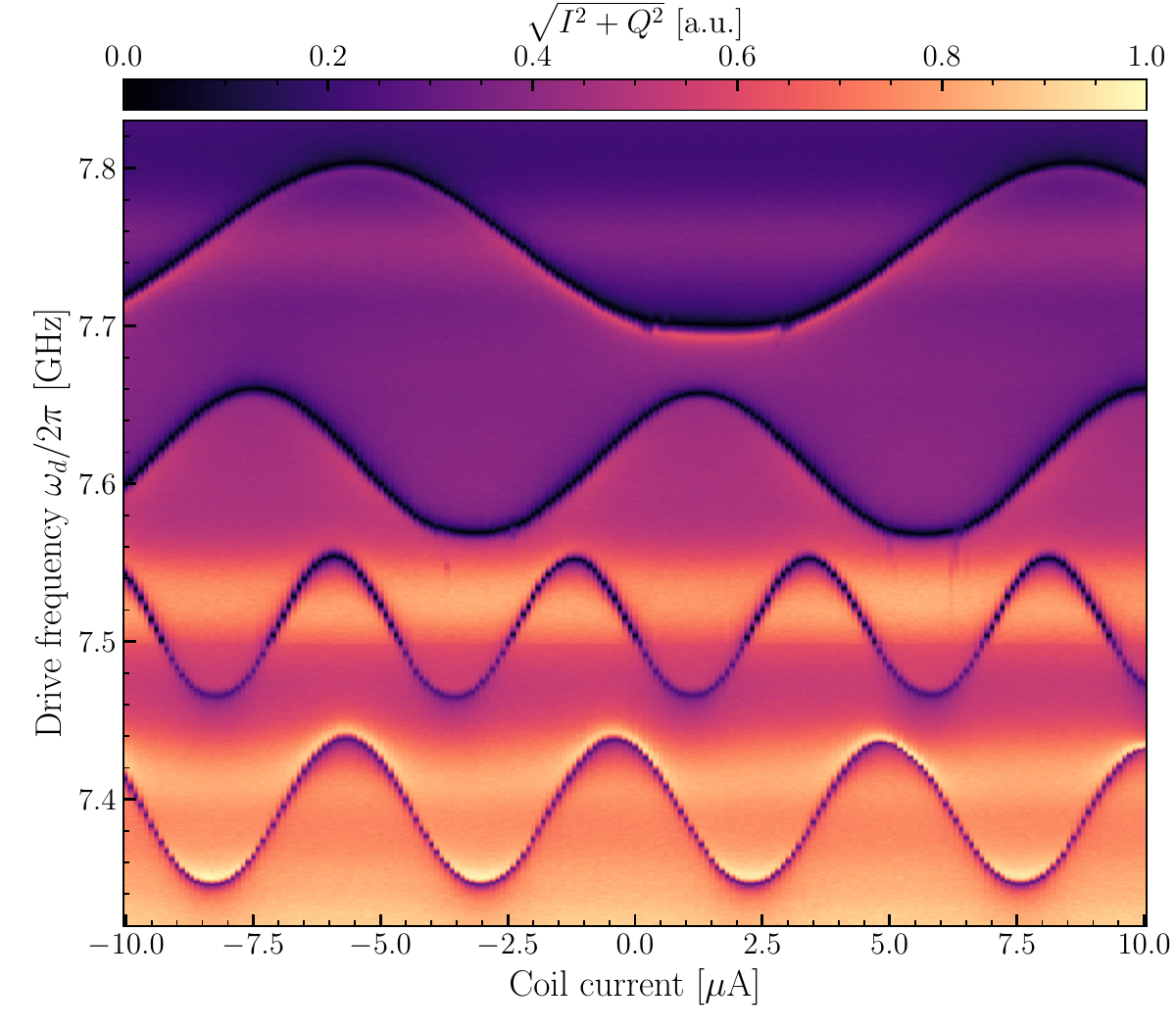}
    \caption{\textbf{Flux spectroscopy of the resonators.} Resonator spectroscopy as a function of coil current, showing fast modulations of the resonator frequency from flux biasing the large loops.}
    \label{sup fig:flux_spectroscopy}
\end{figure}

In addition, we investigate the flux dependence of the circuit. The device contains three flux loops: two large loops defined by the qubit–resonator geometry and the smaller SQUID loop of the transmon. Because the SQUID loop has a much smaller area, its flux periodicity is well separated from that of the large loops, making the two effects easy to distinguish experimentally. The significantly smaller area of the SQUID loop leads to a distinct flux periodicity from that of the large loops, enabling the two effects to be distinguished experimentally. As shown in \cref{sup fig:flux_spectroscopy}, varying the flux threading the large loops modulates both $E_J$ and $E_{Jc}$, leading to rapid variations in the resonator frequency. To tune the SQUID loop independently, we bias the device at integer multiples of the large-loop flux period, where the resonator frequency reaches its maxima.

\begin{figure}[!h]
    \centering
    \includegraphics[width= 1\columnwidth]{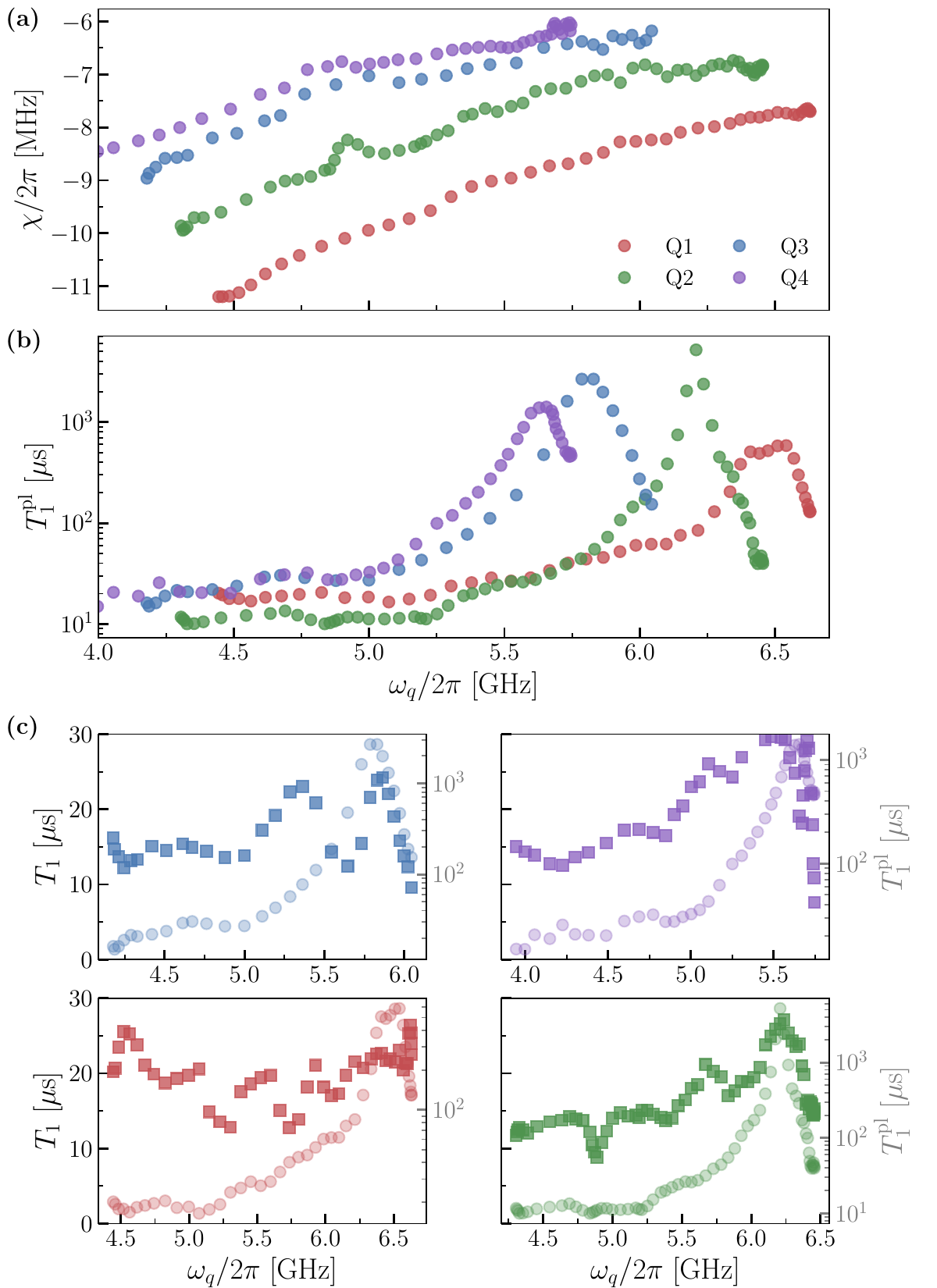}
    \caption{\textbf{Cross-Kerr shifts, Purcell-limited lifetimes, and measured qubit lifetimes} 
    (a) Cross-Kerr shift $\chi$ as a function of qubit frequencies tuned by varying the SQUID flux. The qubit-resonator pairs with a larger $E_{Jc}$ have a stronger cross-Kerr interaction.  
    (b) Purcell-limited relaxation time $T_{1}^{\rm pl}$ versus flux, showing that the intrinsic Purcell filtering (where $T_{1}^{\rm pl}$ reaches maxima) shifts to higher qubit frequencies with increasing $E_{Jc}$ which results in a higher notch frequency.  
    (c) Comparison between $T_{1}^{\rm pl}$ (semi-transparent circles, right axis) and the measured $T_1$ (square, left axis).
    }
    \label{sup fig:flux_response_four_devices}
\end{figure}

\cref{sup fig:flux_response_four_devices} summarizes the cross-Kerr shift $\chi$, measured qubit lifetimes $T_1$, and Purcell-limited qubit lifetimes $T_1^{\rm pl}$ as a function of qubit frequency, tuned by varying the external flux threading the transmon's SQUID loop. Panel (a) shows that the cross-Kerr shift $\chi$ increases with larger $E_{Jc}$ as expected. Panel (b) displays the Purcell-limited relaxation time $T_{1}^{\rm pl}$, obtained with the method outlined in \cref{sec:purcell_protection_ncrits}. All devices exhibit a clear peak at the frequency where the qubit frequency matches the notch-filter frequency set by $E_{Jc}$ and the capacitive coupling strength $J$. Note that the frequency at which Purcell filtering is most effective increases with $E_{Jc}$, since a larger $E_{Jc}$ shifts the notch to higher frequencies. Panel (c) compares the Purcell-limited lifetimes $T_1^{\rm pl}$ with the measured $T_1$ values. For most qubits, the peak in $T_1^{\rm pl}$ coincides with a peak in the measured $T_1$, though the measured lifetimes saturate at lower values due to intrinsic relaxation channels.

\section{Device fabrication and experimental setup}
\label{appendix:device fab and setup}
The device fabrication is entirely carried out at the Center of MicroNano
Technology (CMi) at EPFL. The device is fabricated on a 525-$\mu$m-thick high-resistivity silicon substrate. The fabrication process begins with the definition of alignment markers, which are etched directly into the substrate using photolithography followed by SF$_6$-based dry etching. After stripping the resist, the wafer is cleaned using a piranha solution to remove organic residues. Immediately before metal deposition, the Si native oxide is removed by dipping the wafer in 1\% hydrofluoric acid.  A 150-nm-thick aluminum layer is then deposited by electron-beam evaporation at a rate of 0.2 nm/s. The Al ground plane is patterned via photolithography and wet etching in TechniEtch Al80 for 3 minutes and 30 seconds at room temperature. After stripping the photoresist, the Josephson junctions are fabricated using a Manhattan-style double-angle evaporation process. A bilayer resist (500 nm MMA EL9 and 450 nm PMMA 495k A8) is exposed by electron-beam lithography and developed in a 3:1 MIBK:IPA solution. The aluminum layers forming the junction, which are 85-nm thick and 35-nm thick for the bottom and top layer, respectively, are deposited at 0.5 nm/s under ultra-high vacuum in a Plassys MEB550SL3 system. The tunnel barrier is formed by static oxidation in pure oxygen at 10 Torr for 10 minutes. A second oxidation is performed for capping with the same parameters. To create a galvanic connection between the Josephson junctions and the ground plane, a 200-nm-thick Al patch is deposited by e-beam evaporation at 0.5 nm/s. This patch is patterned via electron-beam lithography and lift-off. Finally, the wafer is diced into $4$ x 7 mm chips and bonded with Al wire to a custom printed circuit board. 

\begin{figure}

    \centering
    \includegraphics[width=0.95 \columnwidth]{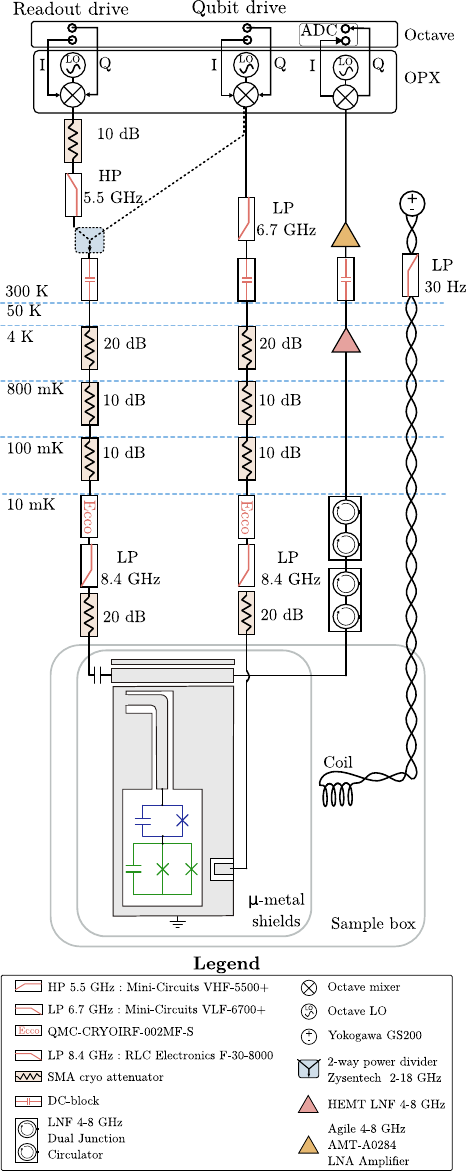}
    \caption{
    \textbf{Schematic of the experimental setup.} 
    The dashed connection, which allows driving the qubit through the feedline, is used only to characterize the device's Purcell protection. The qubit is driven through the dedicated drive line for readout fidelity characterization.}
    \label{sup fig:experimental setup}
\end{figure}

A schematic of the microwave setup is shown in Fig.~\ref{sup fig:experimental setup}. The device is enclosed in a high-purity copper box and thermally anchored to the mixing chamber of a BlueFors dilution refrigerator. Magnetic flux biasing is achieved via a NbTi superconducting coil positioned beneath the chip and driven by a Yokogawa GS200 current source. To suppress any spurious magnetic field, the device is shielded using two layers of high-permeability metal enclosures. A Quantum Machines OPX+ paired with an Octave module is used to generate both qubit and readout drives. 
During the Purcell-limited relaxation time measurements, the qubit and readout drives are combined and sent to the feedline (indicated by dashed lines in the schematic). For all other measurements, the qubit drive is sent via an independent control line. The resonator output passes through two dual-junction circulators (LNF 4-8 GHz Dual Junction Circulator) and is amplified by a 4-8 GHz LNF High-Electron-Mobility Transistor (HEMT) amplifier mounted at the 4K plate. At room temperature, the signal undergoes a last amplification stage with a low-noise amplifier (Agile AMT-A0284), before being demodulated in the Octave and digitized by the OPX+.

\section{Lumped model of junction readout}
\label{appendix:lumped_model_junction_readout}

\begin{figure}
    \centering
    \includegraphics[width=0.65\columnwidth]{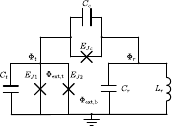}
    \caption{
    \textbf{Lumped-element model of the junction-readout circuit.}
    The transmon qubit is coupled to the resonator through both a Josephson junction and a coupling capacitor.
    }
    \label{sup fig:lumped circuit model}
\end{figure}

We consider a lumped-element model of the junction-readout circuit shown in \cref{sup fig:lumped circuit model} to derive simple analytical relations between the circuit parameters and the Hamiltonian.

The circuit consists of a transmon qubit with capacitance $C_t$ and Josephson energies $E_{J1}$ and $E_{J2}$ which form a SQUID. The transmon is coupled to an LC resonator with capacitance $C_r$ and inductance $L_r$. The transmon is coupled to the resonator with capacitance $C_c$ and a Josephson junction with energy $E_{Jc}$. Furthermore, external flux threads two loops in the circuit. We denote the external flux threading the transmon's SQUID loop as $\Phi_{\rm{ext,t}}$ and the external flux threading the loop that encloses both the transmon and resonator as $\Phi_{\rm{ext,b}}$. To quantize this lumped element circuit we describe the Lagrangian $\mathcal{L} = K - U$ of the system with the generalized flux node variables $\Phi_t$ and $\Phi_r$ that describe the transmon and resonator degrees of freedom, respectively. 

The kinetic energy $K$ of the circuit can be written as, 

\begin{equation}
    K = \frac{C_t}{2} \dot{\Phi}_t^2 + \frac{C_r}{2} \dot{\Phi}_r^2 + \frac{C_c}{2}(\dot{\Phi}_t-\dot{\Phi}_r)^2,
\end{equation}
where the potential energy $U$ is written as, 
\begin{align}
    U &= -E_{J1} \cos{\left(\frac{\Phi_t}{\phi_0}\right)} - E_{J2} \cos{\left(\frac{\Phi_t}{\phi_0} - \frac{\Phi_{\rm{ext, t}}}{\phi_0} \right)} \nonumber \\ &- E_{Jc} \cos{\left(\frac{\Phi_t}{\phi_0} - \frac{\Phi_r}{\phi_0} - \frac{\Phi_{\rm{ext,b}}}{\phi_0} \right)} + \frac{1}{2 L_r} \Phi_r^2.
\end{align}
 To obtain the above expression, we have chosen a spanning tree for the circuit that includes the transmon and resonator shunting capacitances \cite{VoolIJCTA}. Here, $\phi_0 = \Phi_0 / 2\pi = \hbar / 2e$ denotes the reduced flux quantum. 
 The conjugate charge variables are then, 

\begin{align}
    Q_t &= \frac{\partial \mathcal{L}}{\partial \dot{\Phi}_t} = (C_t + C_c) \dot{\Phi}_t - C_c \dot{\Phi}_r, \\ 
    Q_r &= \frac{\partial \mathcal{L}}{\partial \dot{\Phi}_r} = (C_r + C_c) \dot{\Phi}_r - C_t \dot{\Phi}_t.
\end{align}
The Hamiltonian of the circuit can be written as $H = \frac{1}{2} \mathbf{Q}^T \mathbf{C}^{-1} \mathbf{Q} + U$, where 

\begin{align}
    \mathbf{C} = \begin{bmatrix}
C_t + C_c & -C_c \\
-C_c & C_r + C_c
\end{bmatrix},
\end{align}
and $\mathbf{Q} = (Q_t, Q_r)$. This leads to the Hamiltonian, 

\begin{align}
    H &= \frac{Q_t^2}{2 C_{t\Sigma}} -E_{J1} \cos{\left(\frac{\Phi_t}{\phi_0}\right)} -E_{J2} \cos{\left(\frac{\Phi_t}{\phi_0} - \frac{\Phi_{\rm{ext, t}}}{\phi_0} \right)} \nonumber \\ 
    &+ \frac{Q_r^2}{2 C_{r\Sigma}} + \frac{1}{2 L_r} \Phi_r^2 \nonumber \\ 
    &- E_{Jc} \cos{\left(\frac{\Phi_t}{\phi_0} - \frac{\Phi_r}{\phi_0} - \frac{\Phi_{\rm{ext,b}}}{\phi_0} \right)} + \frac{C_c}{\beta} Q_t Q_r,
\end{align}
with $C_{t\Sigma} = C_t + \left( \frac{1}{C_r} + \frac{1}{C_c} \right)^{-1}$, $C_{r\Sigma} = C_t + \left( \frac{1}{C_t} + \frac{1}{C_c} \right)^{-1}$, and $\beta = C_t (C_r + C_c) + C_r C_c$. We promote the classical variables to quantum operators which satisfy the commutation relations $[\hat{\Phi}_i, \hat{Q}_j] = i \hbar \delta_{jk}$ for $j=t$ and $r$ for the transmon and resonator, respectively. Furthermore, we introduce the charge number operator $\hat{n}_j = \hat{Q}_j / 2e$ and phase operator $\hat{\varphi}_j = \hat{\Phi}_j / \phi_0$ to rewrite the Hamiltonian as, 

\begin{align}
    \hat{H} &= 4 E_{Ct} \hat{n}_t^2 - E_J(\Phi_{\rm{ext,t}}) \cos{(\hat{\varphi}_t)} + \omega_r \hat{a}^\dagger \hat{a} \nonumber \\ &+ J \hat{n}_t \hat{n}_r -  E_{Jc} \cos{(\hat{\varphi}_t - \hat{\varphi}_r - \varphi_{\rm{ext,b}})},
\end{align}
with $E_{Ct} = e^2 / 2C_{t\Sigma}$, and $J = 4 e^2 C_c / \beta$. Furthermore, we have introduced the resonator's annihilation and creation operators $\hat{a}$ and $\hat{a}^\dagger$, and the resonator frequency is given by $\omega_r = \sqrt{8 E_{Cr} E_{Lr}}$ with $E_{Cr} = e^2 / 2 C_{r\Sigma}$ and $E_{Lr} = \phi_0^2 / L_r$. The charge and phase operators of the resonator are then defined as $\hat{n}_r = (i/2) (E_{Lr} / 2 E_{Cr})^{1/4} (\hat{a}^\dagger - \hat{a})$ and $\hat{\varphi}_r = (2 E_{Cr} / E_{Lr})^{1/4} (\hat{a}^\dagger + \hat{a})$. Moreover, in the above Hamiltonian we introduce the transmon's effective Josephson energy $E_J(\Phi_{\rm{ext,t}})$ formed by the two Josephson junctions with energies $E_{J1}$ and $E_{J2}$ as, 

\begin{align}
    E_J(\Phi_{\rm{ext,t}}) = E_{J_{\Sigma}} \cos \left( \frac{ \pi \Phi_{\rm{ext,t}}}{\Phi_0} \right) \sqrt{1+d^2 \tan^2 \left( \frac{\pi \Phi_{\rm{ext,t}}}{\Phi_0} \right)},
\end{align}
where $E_{J\Sigma} = E_{J1} + E_{J2}$ and the junction asymmetry is $d = \vert E_{J1} - E_{J2} \vert / E_{J \Sigma}$. Finally, as noted in the main text, we bias $\Phi_{\rm{ext,b}}$ to be at integer flux quanta. We then recover the Hamiltonian in \cref{Eq:Hamiltonian} by setting $E_J(\Phi_{\rm{ext,t}}) = E_J$ as a shorthand notation. 

In practice, our device uses a $\lambda/4$ coplanar-waveguide resonator. The distributed mode is mapped to an effective lumped-element LC oscillator where the fundamental mode has capacitance $C_r = \pi / 4 Z_0\omega_r$ and inductance $4 Z_0 / \pi \omega_r$ \cite{pozar2012microwave} where $Z_0$ is the characteristic impedance of the coplanar waveguide resonator. Given our fitted characteristic impedance of $Z_0 = 54\:\Omega$, this gives an effective mode impedance of $Z_r = 68.8\:\Omega$ for the fundamental mode we use for readout. 

\section{Intrinsic Purcell Protection}
\label{appendix:purcell protection}

\begin{figure}[h]
\centering\includegraphics[width=\linewidth]{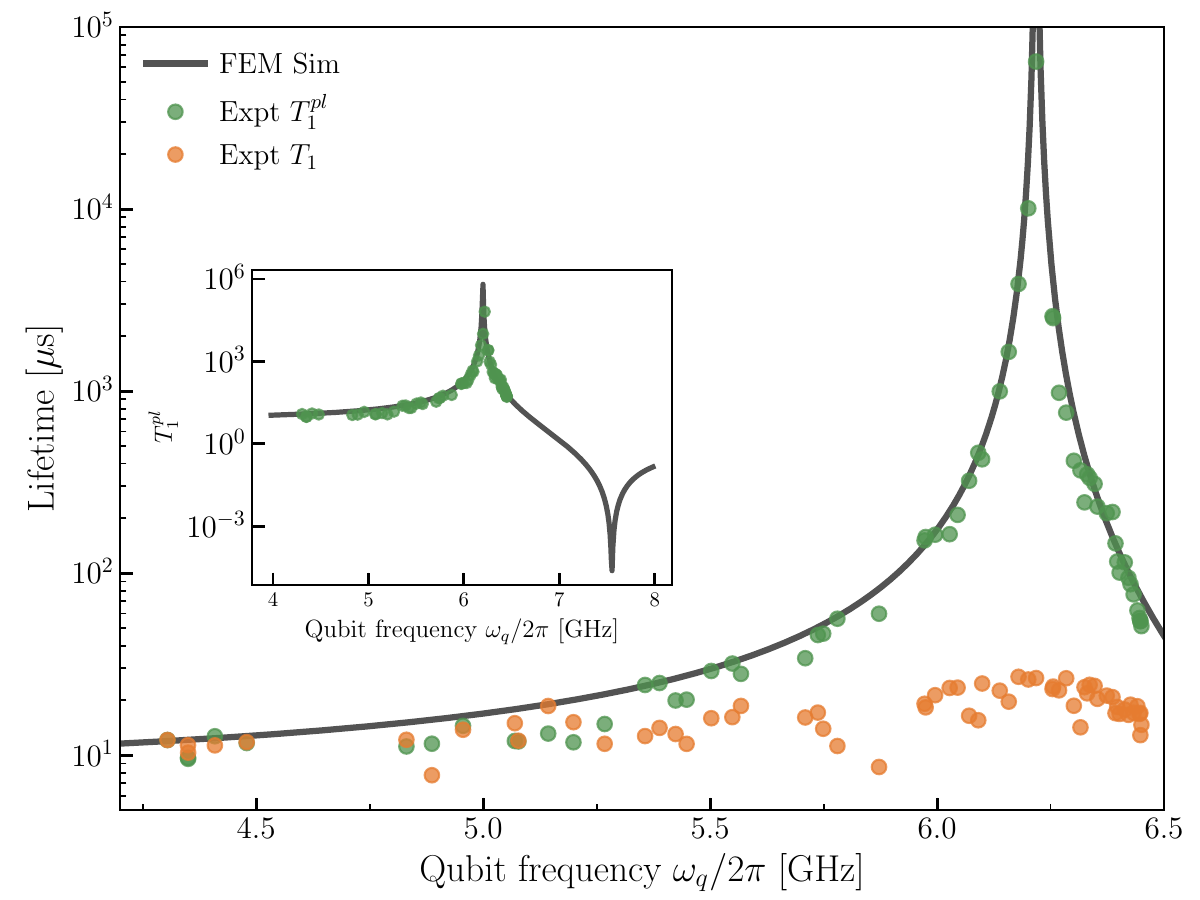}
    \caption{
    \textbf{FEM simulation of the Purcell-limited lifetime.} Purcell-limited relaxation time $T_1^{\rm pl}$ predicted from FEM simulation as a function of qubit frequency  $\omega_q$ (solid black line). Experimentally extracted $T_1^{\rm pl}$ values are shown as green circles, and measured qubit lifetimes $T_1$ as yellow circles. The inset shows the full simulated frequency range and highlights the enhanced decay near the resonator frequency. 
    }
    \label{fig:FEM_purcell_protection}
\end{figure}

The Purcell-limited lifetime can be predicted using finite-element method (FEM) simulation performed in Ansys High Frequency Simulation Software (HFSS) \cite{Yen2025}. The device is modeled as a three-port system comprising the input and output feedline ports, and a third port corresponding to the qubit junction. The qubit relaxation rate to the environment is obtained from the real part of the input admittance $Y_{\rm in}(\omega_q)$ as seen from the qubit port. The Purcell decay rate is then given by \cite{Esteve1986,Neeley2008}:
\begin{equation}
    \frac{1}{T_1^{\mathrm{pl}}} =\frac{\rm{Re[Y_{in}(\omega_q)]}}{C_{t\Sigma}},
\end{equation}
where $C_{t\Sigma}$ is the total transmon capacitance, obtained from Ansys Q3D simulation. The input admittance $Y_{\rm in}$ is extracted from a driven-modal simulation in HFSS. The coupling Josephson junction is modeled as a lumped-element inductor with an inductance of 38.93 nH. This value is treated as a fitting parameter chosen to best match the experimental data. In Fig.~\ref{fig:FEM_purcell_protection}, we compare the Purcell-limited relaxation time predicted by the FEM simulation with the measured data presented in the main text. The excellent agreement between experiment and simulation confirms the intrinsic Purcell-filtering of the circuit.

\section{Semiclassical model for the nonlinear resonator}
\label{appendix:charactherization of the device parameter}
 
As stated in the main text, coupling a transmon and resonator through a Josephson junction introduces a nonlinearity $K$ in the resonator. We model the nonlinear distributed resonator using a single-mode Hamiltonian 
\begin{equation}
    \hat{H}/\hbar=\omega_r \hat{a}^\dagger \hat{a}+(K/2)(\hat{a}^\dagger)^2a^2.
\end{equation}
The equation of motion for the cavity field $\hat{a}$ follows from the quantum Langevin equation, 
\begin{equation}
\label{sup eq: langevin_1}
    \frac{d \hat{a}}{dt} = -i\Delta' \hat{a}-iK\hat{a}^\dagger \hat{a}^2- \frac{\kappa'}{2}\hat{a}+\sqrt{\kappa'}\hat{a}_i ,
\end{equation}
where $\kappa'$ is the coupling rate between the resonator and the feedline, $\Delta' =  \omega_r' -\omega_d$ is the detuning between the drive frequency $\omega_d$ and the resonator frequency $\omega_r'$, and $\hat{a}_i$ is the input field operator. We introduce the primed parameters $\kappa'$, $\omega_r'$, and $\Delta'$ to distinguish the bare resonator parameters from the effective parameters measured experimentally. As shown below, coupling the resonator to the feedline through an input capacitor leads to interference between the incident drive and the field reflected from the output port, resulting in renormalized effective parameters, which we denote $\kappa$, $\omega_r$, and $\Delta$.

\begin{figure}
    \centering
    \includegraphics[width=0.65 \columnwidth]{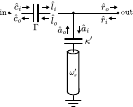}
    \caption{\textbf{Input–output schematic.}
Schematic representation of the resonator and its coupling to the feedline used for the input–output derivation (adapted from Ref.~\cite{Heinsoo_rapid_fidelity}). }
    \label{sup fig:input-output}
\end{figure}

To relate the resonator input field to the drive applied at the device input port, we use the input–output relations for the circuit configuration shown in \cref{sup fig:input-output}. In this configuration, the resonator is coupled to a feedline engineered to preferentially decay toward the output port by introducing asymmetry via the input capacitance $C_{\rm in}$. The corresponding input-output relations for this circuit were derived in Ref.~\cite{Heinsoo_rapid_fidelity}. For completeness, we briefly recall the main results below.

The resonator input field $\hat{a}_i$ is related to its output field $\hat{a}_0$ through 
\begin{equation}
 \hat{a}_o=\hat{a}_i - \sqrt{\kappa'} \hat{a}.
\end{equation}
The scattering matrix of the T-junction connecting the these resonator fields to the left ($\hat{l}$) and right ($\hat{r}$) ports of the device is described by the following relations,
\begin{equation}
  \hat{l}_o = - \frac{1}{3} \hat{l}_i +\frac{2}{3}\hat{r}_i + \frac{2}{3}\hat{a}_o,
\end{equation}
\begin{equation}
\hat{r}_o = \frac{2}{3}\hat{l}_i - \frac{1}{3}\hat{r}_i + \frac{2}{3}\hat{a}_0,
\end{equation}
\begin{equation}
    \hat{a}_i=\frac{2}{3}\hat{l}_i+\frac{2}{3}\hat{r}_i-\frac{1}{3}\hat{a}_o,
\end{equation}
where $\hat{l}_i$, $\hat{r}_i$ ($\hat{l}_o$, $\hat{r}_o$) denote the input (output) field modes propagating in the left and right directions of the feedline, respectively. The effect of the input capacitor is introduced via a reflection coefficient $\Gamma(\omega)=1/(1+2i\omega Z_0 C_{\mathrm{in}})$, which relates the input ($\hat{c}_i$) and output ($\hat{c}_o$) signals at the left port to the feedline modes through the following expressions:
\begin{equation}
    \hat{c}_o = (1-\Gamma)\hat{l}_o + \Gamma \hat{c}_i,
\end{equation}
\begin{equation}
    \hat{l}_i=(1-\Gamma)\hat{c}_i + \Gamma \hat{l}_o.
\end{equation}
Using the above expressions, we obtain the input-output relations for both the mirror side and the right port of the feedline
\begin{align}
    \hat{c}_o &= (1-\Gamma)\hat{r}_i + \Gamma \hat{c}_i -(1-\Gamma) \frac{\sqrt{\kappa'}}{2} \hat{a}, \\ 
    \hat{r}_o &= (1-\Gamma)\hat{c}_i + \Gamma \hat{r}_i -(1+\Gamma) \frac{\sqrt{\kappa'}}{2}\hat{a},
\end{align}
Substituting these relations into the Langevin equation yields an equation of motion of the same form as \cref{sup eq: langevin_1}, but with a renormalized resonator frequency and linewidth, 
\begin{align}
\dot{\hat a} &= -i\Delta \hat a - iK\hat a^\dagger \hat a^2  - \frac{\kappa}{2}\hat a \nonumber \\
&\quad + \frac{\sqrt{\kappa'}}{2}\big[(1-\Gamma)\hat c_i + (1+\Gamma)\hat r_i\big].
\end{align}
The corresponding effective parameters are
\begin{equation}
\label{sup eq:effective kappa}
    \kappa=\frac{\kappa'}{2}[1+\rm{Re(\Gamma)}], 
\end{equation}
and
\begin{equation}
\label{sup eq:effective omega}
     \omega_r=\omega_r'+\kappa'\frac{\rm Im({\Gamma})}{4},
\end{equation}
such that the detuning in the equation of motion is defined with respect to the effective resonance frequency as $\Delta=\omega_r-\omega_d$. 

In our configuration we drive only through the left port ($\hat{r}_i=0$). In the semiclassical steady-state regime, we assume that the cavity and input fields are in coherent states and replace the corresponding operators by their expectation values, 
$\alpha = \langle \hat a \rangle$, $c_{i} = \langle \hat c_i \rangle$. Setting $\dot \alpha = 0$, we obtain
\begin{equation}
\label{sup eq: steady state}
    i\Delta \alpha + iK |\alpha|^2 \alpha +\frac{\kappa}{2} \alpha = F,
\end{equation}
with $F = \frac{\sqrt{\kappa'}}{2}(1 - \Gamma)c_i.$ The average intracavity photon number $n=|\alpha|^2$ can be obtained by multiplying \cref{sup eq: steady state} by its complex conjugate \cite{Eichler_2014_josephson}, giving 
\begin{equation}
\label{sup eq: Kerr photon number}
    |F|^2=\left(\Delta^2+\frac{\kappa^2}{4} \right) n+2K\Delta n^2  +K^2n^3,
\end{equation}
where $|F|^2=\frac{\kappa'}{4}|1-\Gamma|^2 \langle \hat{c}_i^\dagger \hat{c_i} \rangle = \frac{\kappa'}{4}|1-\Gamma|^2 \frac{P}{A \hbar \omega}$.

Solving \cref{sup eq: Kerr photon number} for the steady-state intracavity photon number $n$ gives a nonlinear response that can exhibit multiple real solutions depending on the system parameters. A bifurcation occurs when the system transitions from a single stable solution to three solutions (two stable and one unstable). This critical point is characterized by the conditions $\partial \Delta / \partial n=0$ and $\partial^2 \Delta / \partial n^2=0$. For a purely Kerr nonlinear system, these conditions give the well-known result $\Delta_{\rm bif}=-{\rm sgn} (K)\sqrt{3}\kappa/2$ and a corresponding photon number $n_{\rm bif}=\frac{\kappa}{\sqrt{3}K}$ \cite{Eichler_2014_josephson}. 

In the main text, we use \cref{sup eq: Kerr photon number} to fit the measured steady-state intracavity photon number as a function of drive frequency and drive amplitude. The only free fitting parameter is the effective drive amplitude 
$F$. The self-Kerr is fixed to the state-averaged value $K/2\pi=\frac{1}{2}(K_g/2\pi+K_e/2\pi)=-0.963$ MHz, where $K_g$ and $K_e$ are the Kerr coefficients obtained from the Hamiltonian \cref{Eq:Hamiltonian} using the previously extracted parameters, with the qubit initialized in $\ket{g}$ and $\ket{e}$, respectively. The remaining parameters are set to their measured values: $\kappa/2\pi=10.6$ MHz, $\chi/2\pi=-7$ MHz and $\omega_r/2\pi=7.659$ GHz.

\begin{figure}
    \centering
    \includegraphics[width= \columnwidth]{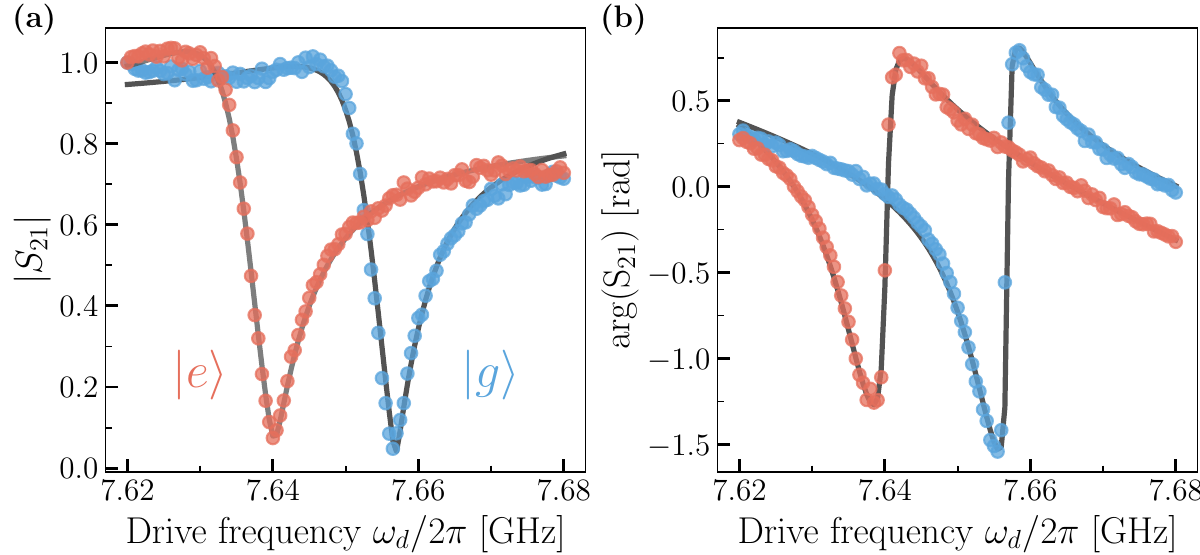}
    \caption{
    \textbf{$\mathbf{S_{21}}$ measurement}. Measured transmission coefficient $S_{21}$ with the qubit prepared in $\ket{g}$ (blue circles) and $\ket{e}$ (red circles). Solid lines are fits to \cref{sup eq: s21 fit}. 
    }
    \label{sup fig:cross-kerr measurement}
\end{figure}

Experimentally, we can also characterize the system by measuring the scattering coefficient $S_{21}=\frac{\expval{\hat{r}_o}}{\expval{c_i}}|_{\expval{\hat{r}_i}=0}$. Below, we consider the regime of very weak driving, in which the nonlinear terms are negligible and the scattering coefficient is given by
\begin{equation}
\label{sup eq:S21 theory}
    S_{21} = (1 - \Gamma) \left[ 1 - \frac{(1 + \Gamma)}{1 + \operatorname{Re}(\Gamma)}  \frac{\kappa}{2i\Delta + \kappa} \right].
\end{equation}
To perform a direct fit of the experimental data with \cref{sup eq:S21 theory}, we include an additional correction prefactor that accounts for frequency-dependent attenuation or gain in the measurement line and for phase accumulation due to finite propagation speed along the cables. Specifically, we multiply \cref{sup eq:S21 theory} by $(A_0+A_\omega \omega)e^{-i(\alpha+\tau \omega)}$, where $A_0$ and $A_{\omega}$ describe a linear dependence of the amplitude response, $\omega$ is the probe frequency, $\alpha$ is a constant phase offset, and $\tau$ is the electronic delay time constant. Hence, the scattering coefficient can be expressed as,
\begin{align}
\label{sup eq: s21 fit}
S_{21} &= (A_0 + A_\omega \omega)e^{-i(\alpha+\tau \omega)}(1 - \Gamma) \nonumber \\
&\quad \times \left[ 1 - \frac{1 + \Gamma}{1 + \operatorname{Re}(\Gamma)}
        \frac{\kappa}{2 i \Delta + \kappa} \right].
\end{align}
\cref{sup fig:cross-kerr measurement} shows an example $S_{21}$ parameters for weak drive with the qubit in the $\ket{g}$ (blue) and $\ket{e}$ (red). We find excellent agreement between the measured transmission coefficients $S_{21}$ and the numerical fits. 

\section{Stark-shift measurement}
\label{Appendix:Calibration}

We use the ac Stark shift to calibrate the resonator parameters, such as the coupling rate $\kappa$, the steady-state intracavity photon number, and the photon number during the readout pulse, following Ref.~\cite{Sank2025}.

\begin{figure}[!h]
    \centering \includegraphics[width= 1\columnwidth]{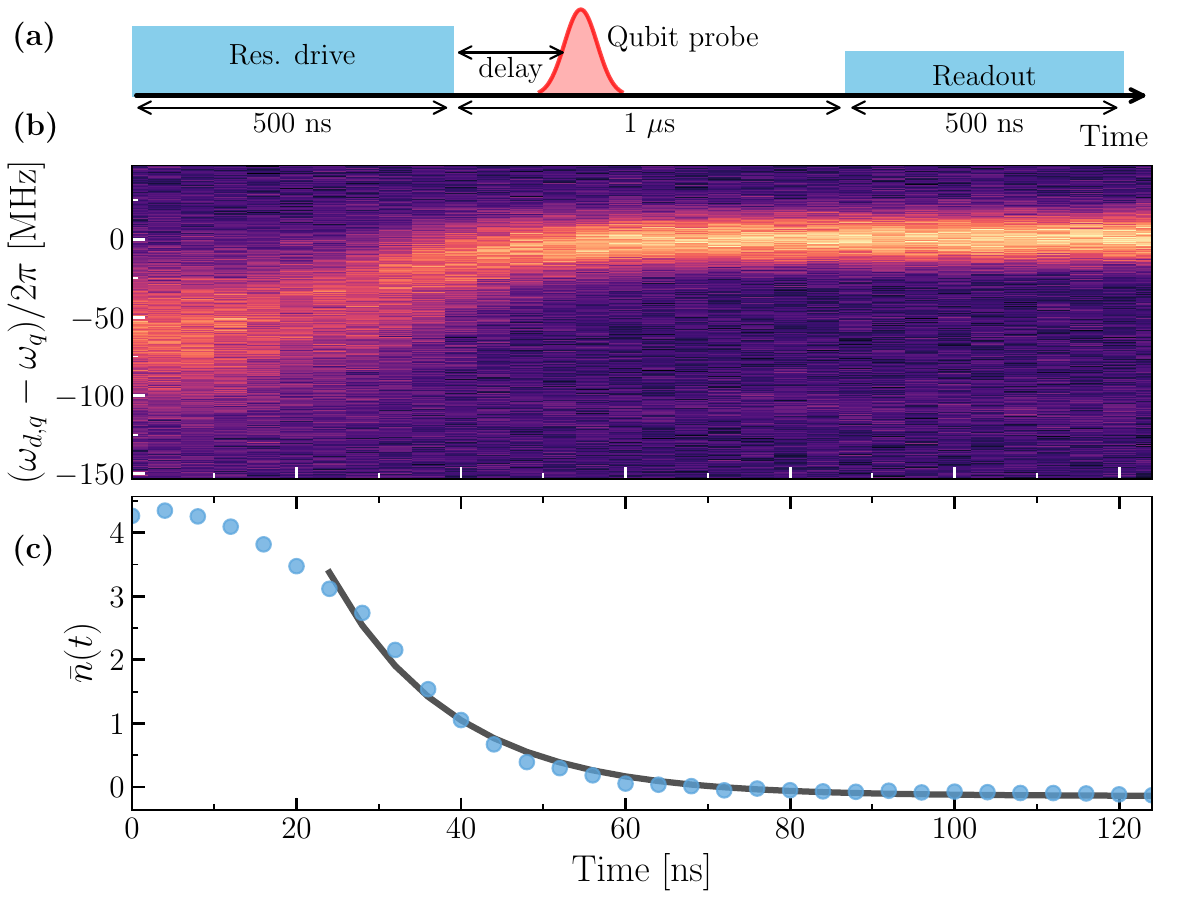}
    \caption{\textbf{Resonator decay rate measurement.} 
    (a) Pulse sequence used for calibrating the resonator decay rate $\kappa$. The red trace corresponds to the qubit probe pulse, while the blue boxes indicate the drives on the resonator. 
    (b) Probability of measuring the qubit in $\ket{e}$ as a function of time after the resonator drive is switched off.
    (c) Intracavity photon number extracted from the Stark shift (circles), with the exponential fit (solid line) used to determine the resonator decay rate $\kappa$.
    }
    \label{sup fig:ac-stark-time}
\end{figure}

To determine $\kappa$, we use the pulse sequence sketched in Fig.~\ref{sup fig:ac-stark-time}(a). The qubit is initialized in $\ket{g}$ while the resonator is driven for $500$\,ns, a duration much longer than $1/\kappa$, to ensure a steady-state resonator photon population $\bar{n}$. The resonator population shifts the qubit frequency by $\delta_q \approx 2\chi \bar{n}$. After turning off the resonator drive, a 48-ns Gaussian-envelope qubit probe pulse of variable frequency is applied with increasing delay times. By measuring the qubit excitation probability as a function of delay, we extract the Stark-shifted qubit frequency and determine the photon population as the resonator decays. The photon number follows an exponential decay, as shown in \cref{sup fig:ac-stark-time}(b) and (c). From the exponential fit, we obtain a resonator linewidth of $\kappa/2\pi=10.6$ MHz.

\begin{figure}[!h]
    \centering
    \includegraphics[width= 1\columnwidth]{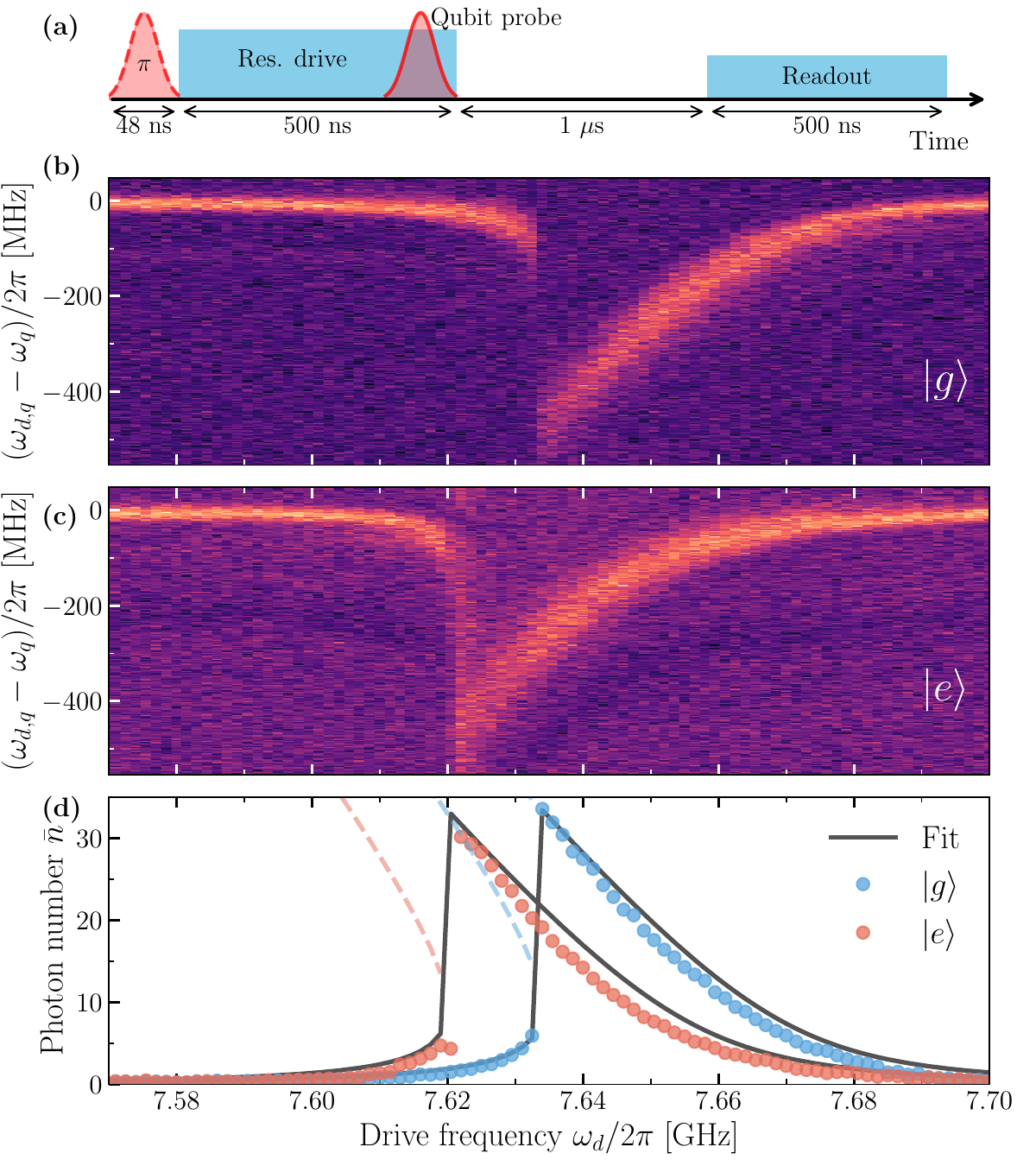}    
    \caption{
    \textbf{Intracavity photon number calibration}. (a) Pulse sequence. The red trace corresponds to the qubit probe pulse, and the blue boxes indicate drives on the resonator. The dashed red trace indicates the qubit preparation pulse that is present (absent) when preparing the qubit in the state $\ket{e}$ ($\ket{g}$). 
    (b)-(c) Probability of measuring the qubit in state $\ket{e}$ ($\ket{g}$) when the qubit was prepared in state $\ket{g}$ ($\ket{e}$) for varying resonator drive frequencies $\omega_d$. 
    (d) Extracted intracavity photon number vs resonator drive frequency $\omega_d$ for the qubit prepared in $\ket{g}$ or $\ket{e}$. The blue (red) points are the extracted measurements for $\ket{g}$ ($\ket{e}$). The solid lines are fits to \cref{sup eq: Kerr photon number}. }
    \label{sup fig:photon number calibration}
\end{figure}

The steady-state intracavity photon number is calibrated with a similar sequence, shown in Fig.~\ref{sup fig:photon number calibration}(a). The qubit is prepared in either $\ket{0}$ or $\ket{1}$, and the resonator is driven at frequency $\omega_d$ for $500$\,ns to reach steady state. The resulting resonator occupation shifts the qubit frequency. A probe pulse at frequency $\omega_{d,q}$ is then applied to the qubit. The qubit is flipped when $\omega_{d,q}$ is resonant with the Stark-shifted qubit frequency. After the resonator drive is turned off and the resonator is allowed to deplete, the qubit state is measured through the same resonator. By sweeping both $\omega_d$ and $\omega_{d,q}$, we obtain a two-dimensional excitation map, as shown in \cref{sup fig:photon number calibration}(b) and (c). From these measurements, the intracavity photon number is extracted and fitted using \cref{sup eq: Kerr photon number}, following the procedure described in Appendix~\ref{appendix:charactherization of the device parameter}.

\begin{figure}[!h]
    \centering
\includegraphics[width= 1\columnwidth]{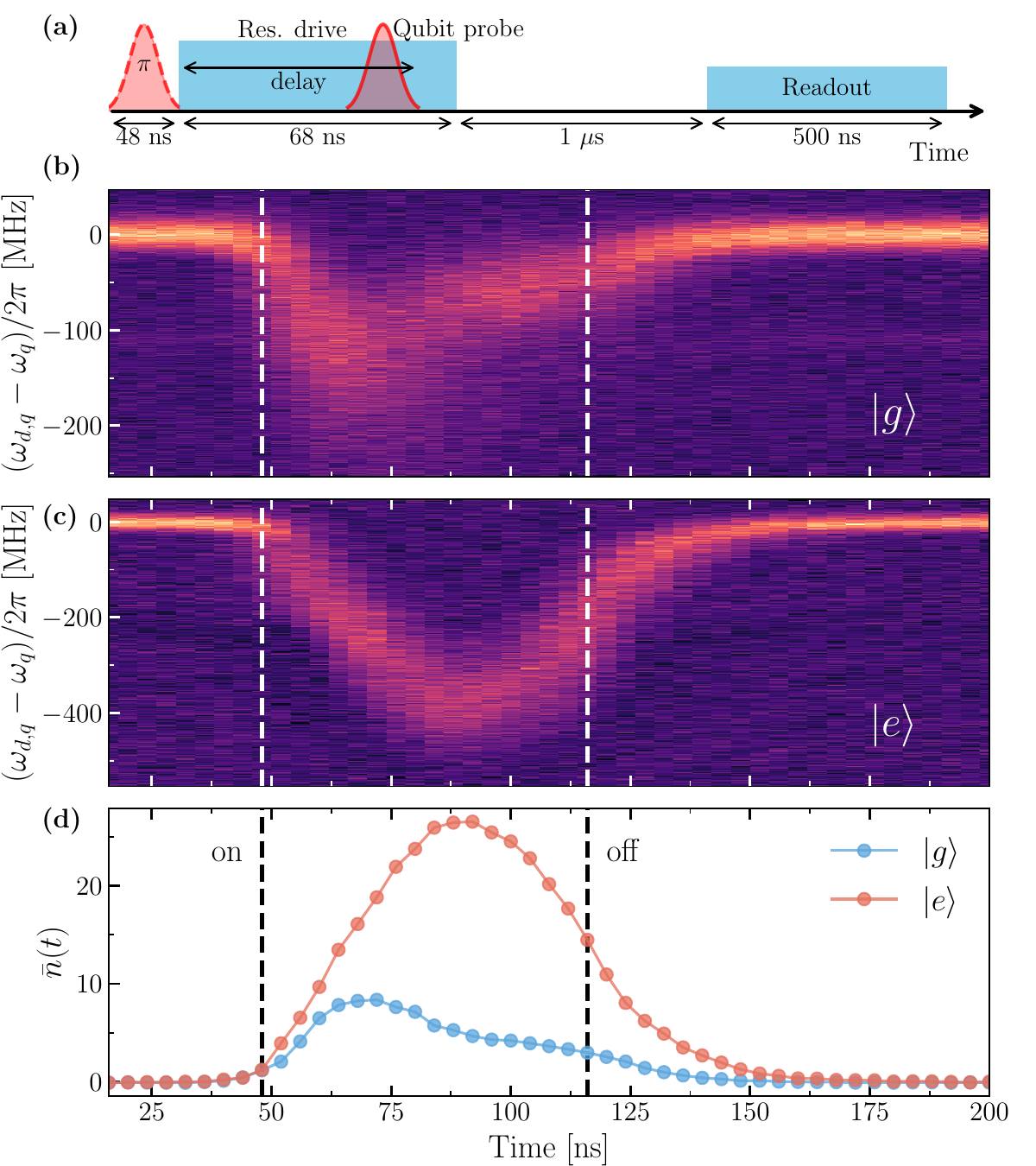}
    \caption{\textbf{Intracavity photon number during the readout pulse.} 
    (a) Pulse sequence. The blue boxes represent resonator drive pulses, and the red curve denotes the qubit probe pulse. The dashed red line at the beginning indicates the optional preparation pulse, absent (present) for initialization in $\ket{0}$ ($\ket{1}$). 
    (b–c) Probability of measuring the qubit in the state opposite to the one prepared, for initialization in $\ket{g}$ and $\ket{e}$, respectively, as a function of time during the resonator drive. 
    (d) Extracted intracavity photon number versus time for the qubit prepared in $\ket{g}$ (blue) and $\ket{e}$ (red).}
    \label{sup fig:intra_cavity_photon_number}
\end{figure}

Finally, the intracavity photon number during the readout pulse is obtained using the ac Stark shift, following the same protocol used to calibrate the resonator decay rate (\cref{sup fig:photon number calibration}). The qubit is first prepared in either $\ket{g}$ or $\ket{e}$ using a $\pi$ pulse and is then continuously probed during the resonator drive. This allows us to monitor the photon population in the resonator throughout the readout drive. The result of this measurement are shown in \cref{sup fig:intra_cavity_photon_number}(b-d). We note that when the qubit is prepared in $\ket{e}$, the intracavity photon number begins to decay before the end of the drive pulse. This occurs because the resonator dynamics do not follow a simple exponential ring-up, but instead exhibit oscillatory transients.

\section{Quantum efficiency}
\label{Appendix_Quantum_efficiency}

\begin{figure}[!h]
    \centering \includegraphics[width= 1\columnwidth]{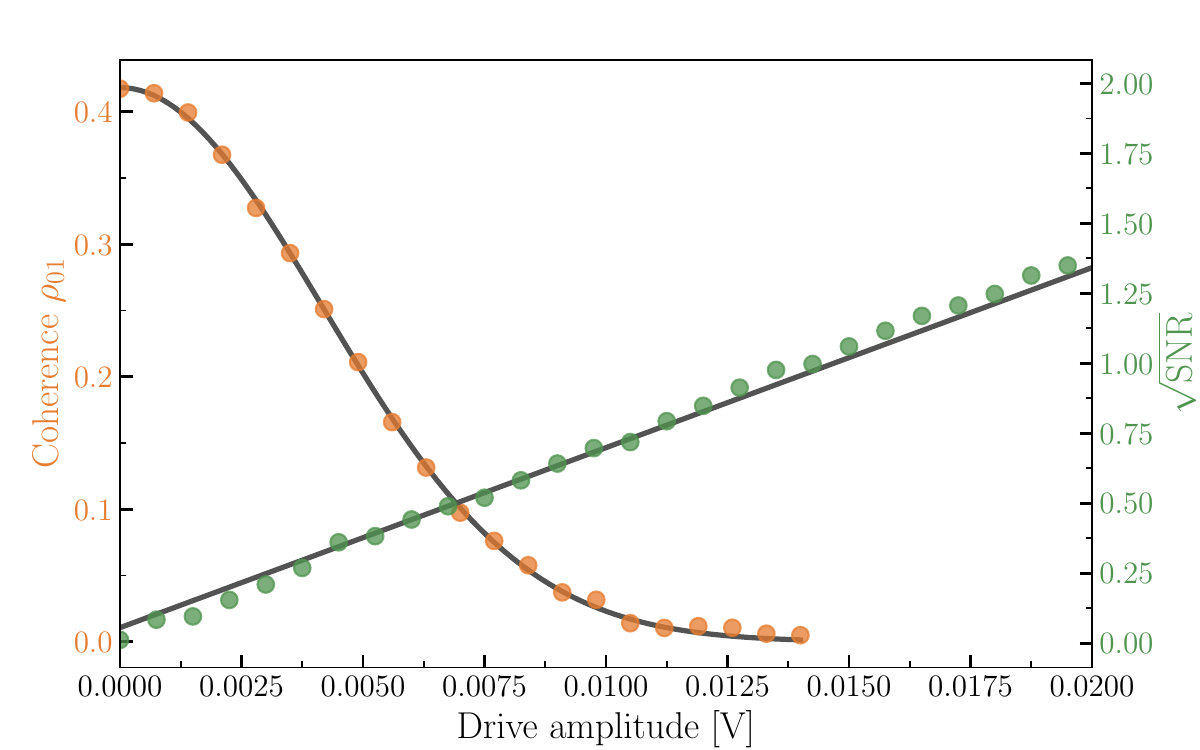}
    \caption{\textbf{Quantum efficiency measurement.} Measurement of the quantum efficiency in the linear regime. Orange circles: Ramsey-fringe contrast versus measurement amplitude, with the solid line corresponding to the fit to the dephasing model 
    $ |\rho_{ge}(\epsilon)| = |\rho_{ge}(0)| e^{-\epsilon^2/(2\sigma_m^2)} $. 
    Green circles: $
    \sqrt{{\rm SNR}}$ versus measurement amplitude, with the solid line showing the linear fit used to extract the proportionality constant $c$.}
    \label{sup fig:quantum_efficienc}
\end{figure}

We determine the quantum efficiency $\eta$ using the method of Ref.~\cite{Bultink2018}, by comparing the measured SNR to the measurement-induced dephasing rate of the qubit. For a weak measurement in the linear regime, the SNR of the integrated readout signal is defined as 
\begin{align}
    {\rm SNR}=\left| \frac{\mu_e-\mu_g}{(\sigma_g+\sigma_e)/2} \right|^2,
\end{align} 
where $\mu_{e/g}$ and $\sigma_{e/g}$ are the means and standard deviations of the distributions obtained when preparing the qubit in $\ket{e}$ or $\ket{g}$, respectively. In this regime, the SNR scales quadratically with the drive amplitude $\epsilon$, ${\rm SNR}=c^2 \epsilon^2$ , where $c$ is a proportionality constant obtained by fitting $\sqrt{ \rm SNR}$ versus $\epsilon$ (green circles in Fig.~\ref{sup fig:quantum_efficienc}). The same measurement drive induces dephasing of the qubit. The coherence decays as $|\rho_{ge}(\epsilon)|=|\rho_{ge}(0)|e^{-\beta_m}$ with $\beta_m=\frac{\epsilon^2}{2 \sigma_m^2}$, where $\sigma_m$ is obtained by fitting the decay of Ramsey fringes contrasts versus measurement amplitude (orange circles in Fig.~\ref{sup fig:quantum_efficienc}). The quantum efficiency is then obtained from 
\begin{equation}
    \eta = \frac{{\rm SNR}}{4 \beta_m}  = \frac{\sigma_m^2 c^2}{2}.
\end{equation}
With this definition, the quantum efficiency satisfies $0\leq \eta \leq 1/2$.
Using the method described above, we measure an efficiency of $\eta \approx 4$\,\%. As noted in the main text, the low efficiency results from using a $4$K HEMT as a first stage amplifier. Future generations of the device can benefit from using near-quantum-limited amplifiers to further improve the readout performance. 

\section{Discussion on fidelity and QNDness error}
\label{Appendix_QNDness_error}
\begin{figure}
    \centering
 \includegraphics[width= 1\columnwidth]{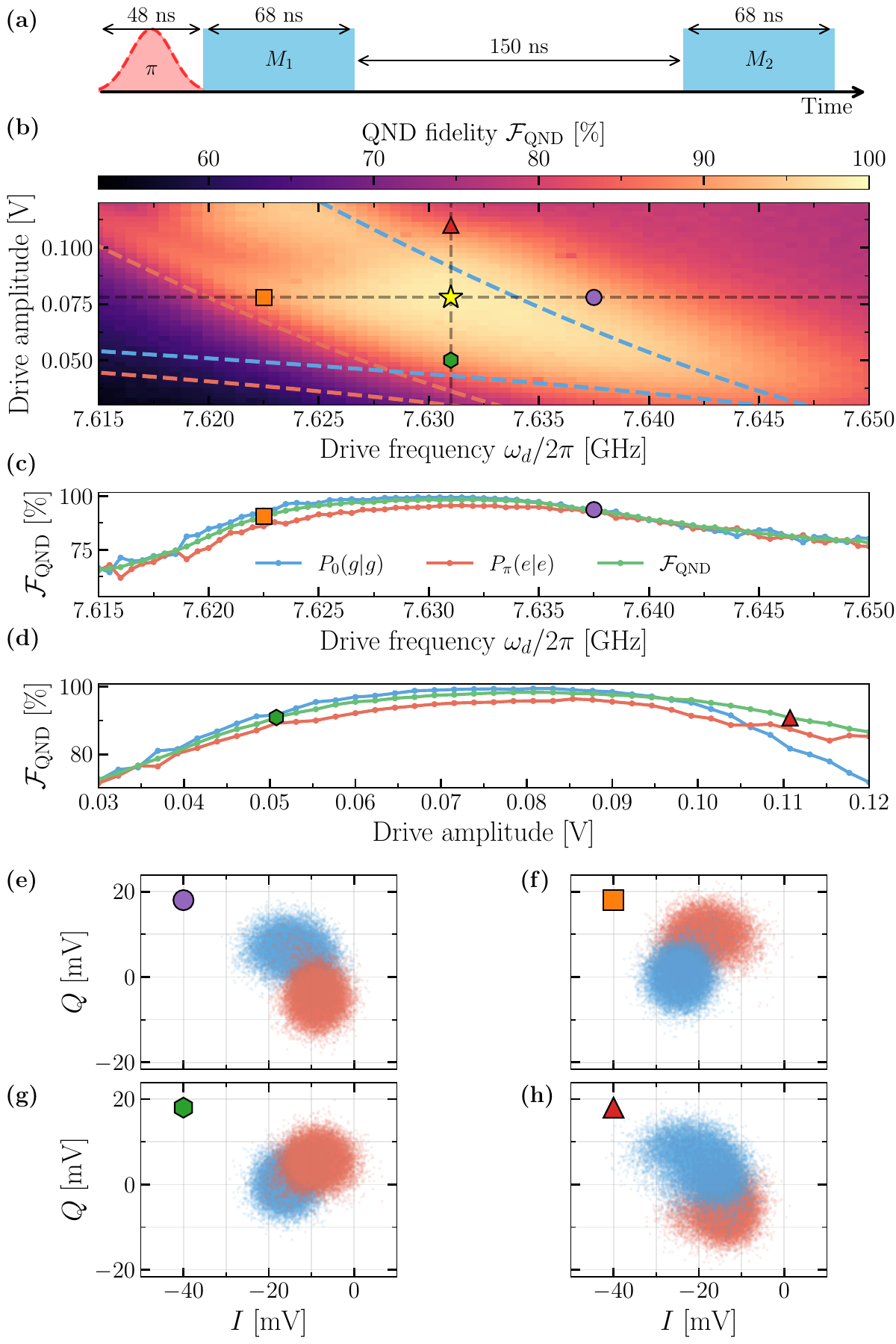}
    \caption{
    \textbf{Bistability and QND fidelity.} (a) Pulse sequence used to measure QND fidelity. See \cref{Fig:QND_main} and \cref{sec:fidelity_and_QNDness} for details. (b) QND fidelity as a function of drive frequency $\omega_d$ and amplitude. The star marker indicates the operating point used to characterize the readout fidelity in the main text. The colored dashed lines indicate the bistable regions of the resonator for when the qubit is prepared in $\ket{g}$ (blue) and $\ket{e}$ (red), while the black dashed lines refer to the cuts shown in panels (c) and (d). 
    (c) QND fidelity as a function of drive frequency $\omega_d$. (d) Same as (c) but as a function of drive amplitude. (e)-(h) IQ distributions of $5\times10^4$ shots of the first measurement outcomes taken at the colored marker points in (b) (orange square, red triangle, purple circle, and green hexagon).
    }
    \label{sup fig:QND fidelity_extra}
\end{figure}

In this section, we further examine the potential limitations of bifurcation-based readout on the measurement fidelity and QND performance. \cref{sup fig:QND fidelity_extra} supplements the results shown in \cref{Fig:QND_main}. For reference, \cref{sup fig:QND fidelity_extra}(a) and (b) show the pulse sequence and readout fidelity map from \cref{Fig:QND_main}. \cref{sup fig:QND fidelity_extra}(c) and (d) show line cuts along the dashed traces in \cref{sup fig:QND fidelity_extra}(b), illustrating the dependence of the QND fidelity on the drive frequency $\omega_d$ and drive amplitude. We observe that the fidelity drops abruptly near the boundaries of the bistable region which are indicated by the colored dashed contours in \cref{sup fig:QND fidelity_extra}(b). 

\begin{figure}
    \centering
    \includegraphics[width= 1\columnwidth]{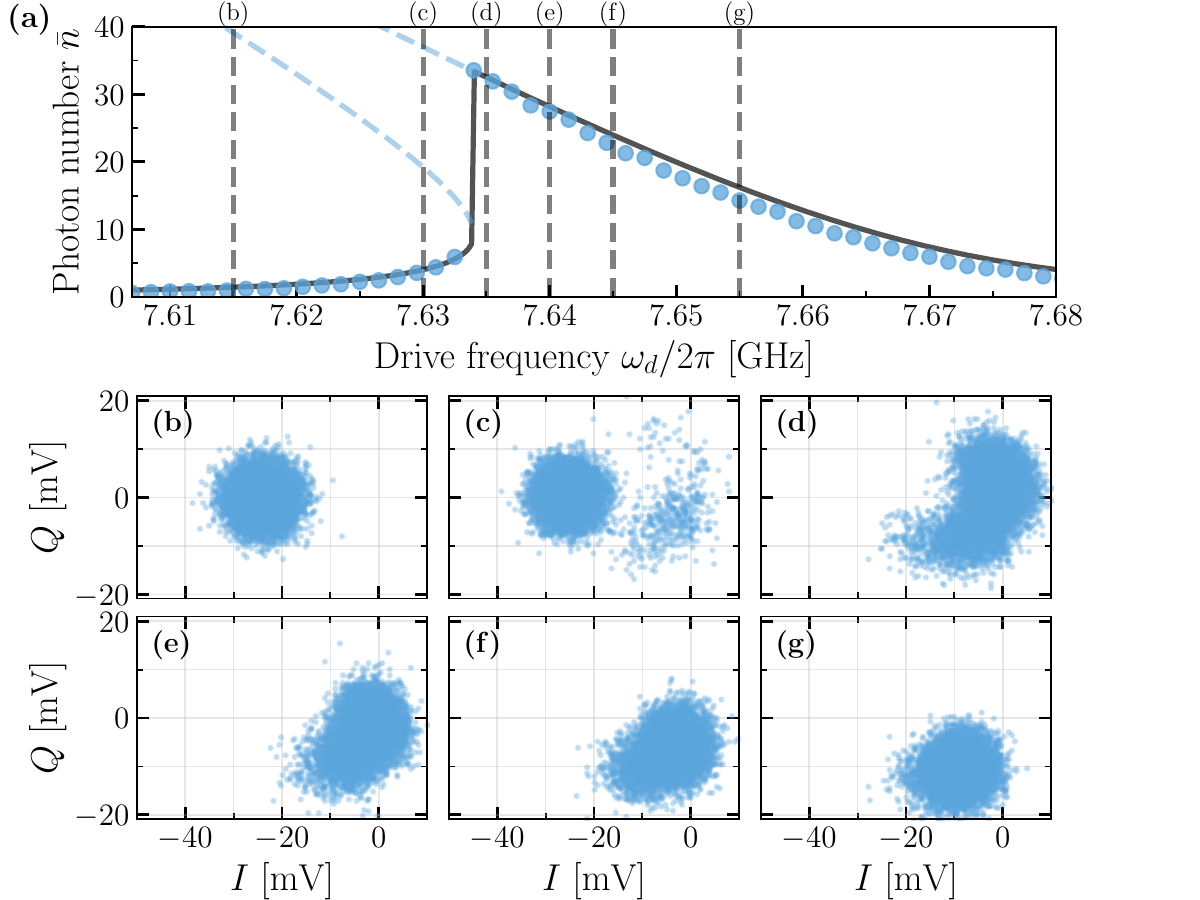}
    \caption{
    \textbf{IQ distributions near the bifurcation transition.}  (a) Steady-state photon number versus drive frequency $\omega_d$ for qubit prepared in $\ket{g}$. Circular markers indicate measured data while the solid line is a fit to \cref{sup eq: Kerr photon number}. The light dashed lines indicate the bistable regions. 
    (b)-(g) IQ distributions of $5 \times 10^4$ readout events at the drive frequencies marked in (a). Away from the bifurcation point, the distributions are well described by Gaussians, indicating the absence of switching events. In contrast, near the bifurcation point frequent switching between the low- and high-photon-number resonator states is observed, producing spurious points in the IQ plane, see (c)–(d).
    }
    \label{sup fig:photon_number_IQ_freq}
\end{figure}

The origin of this degradation is visible in the IQ distributions shown in \cref{sup fig:QND fidelity_extra}(e–h). Near the upper boundary of the bistable region for the resonator when the qubit is prepared in the ground state (corresponding to the region close to the red triangle and purple circle in \cref{sup fig:QND fidelity_extra}(b)), we observe trajectories in which the resonator switches from the low- to the high-photon-number state during the measurement. Similarly, near the upper boundary of the bistable region for the resonator when the qubit is prepared in the excited state (the region close to the orange square in \cref{sup fig:QND fidelity_extra}(b)), trajectories are observed in which the resonator switches from the high- to the low-photon-number state during measurement. Such switching events introduce additional errors that degrade both the assignment fidelity and the QND fidelity. This effect is further illustrated in \cref{sup fig:photon_number_IQ_freq}, which shows the IQ distributions as a function of drive frequency for the qubit prepared in $\ket{g}$, at a fixed drive amplitude corresponding to the horizontal dashed line in \cref{sup fig:QND fidelity_extra}(b). When the resonator is driven at frequencies close to the transition between low- and high–photon-number states, numerous switching events give rise to spurious points in the IQ plane, as shown in \cref{sup fig:photon_number_IQ_freq}(c) and (d). This behavior is a known effect of bifurcation-based readout \cite{Mallet2009,Vijay2009,Andersen2020}. 

\begin{figure}[!h]
    \centering
    \includegraphics[width= 1\columnwidth]{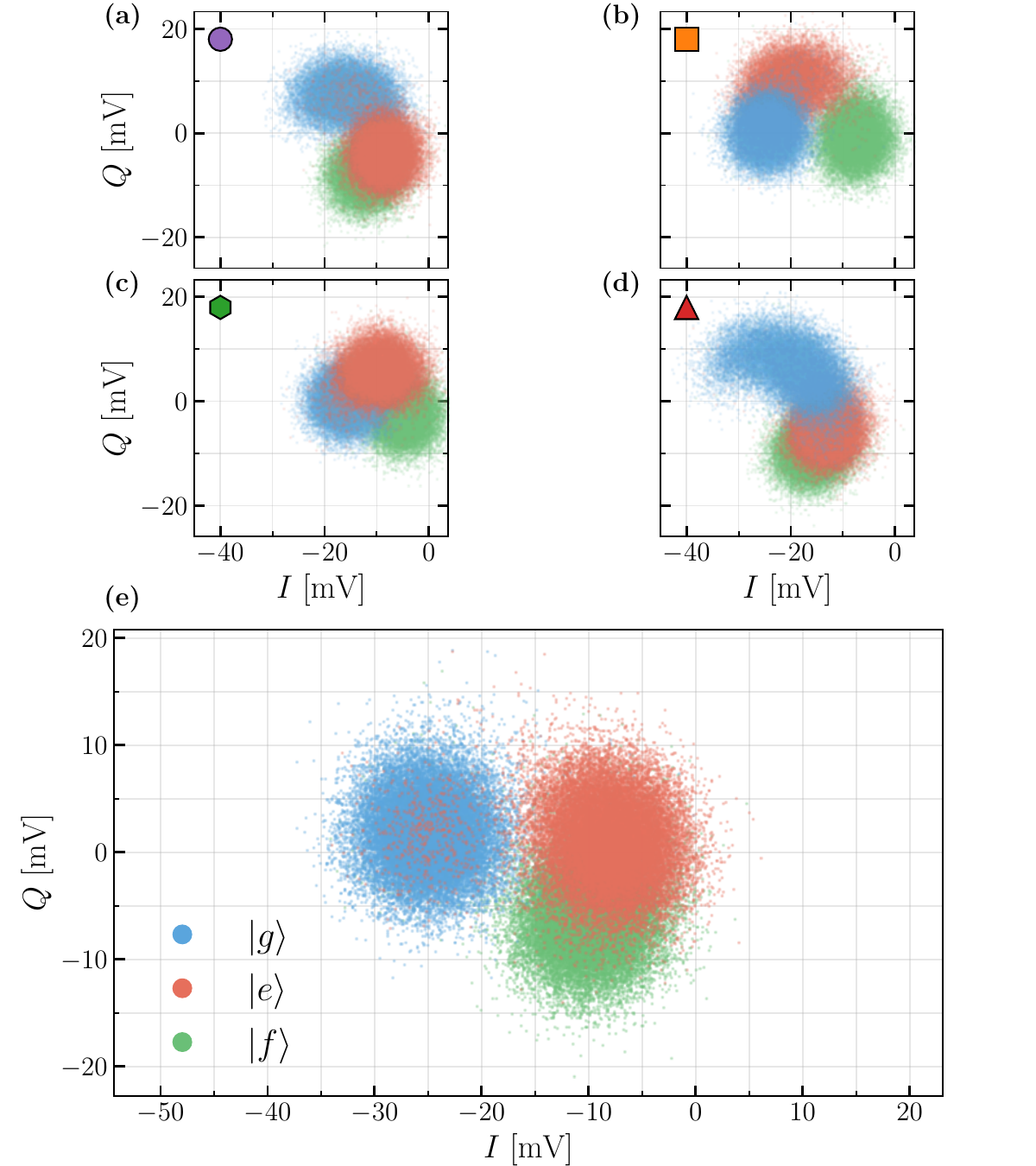}
    \caption{
    \textbf{Second excited state distribution.} 
    (a–d) IQ distributions of $10^4$ measurements for the qubit prepared in $\ket{g}$ (blue), $\ket{e}$ (red), and $\ket{f}$ (green), taken at the four operating points indicated in \cref{sup fig:QND fidelity_extra}(a). The absence of overlap between the $\ket{f}$ state distribution and the switching region suggests that the observed spurious points arise from resonator dynamics, not from leakage to $\ket{f}$. (e) IQ distribution of $10^4$ measurements at the star-marked operating point of \cref{sup fig:QND fidelity_extra}(a), where switching events are still visible in the lower part of the distribution.
    }
    \label{sup fig:Second_excited_state}
\end{figure}

Nonetheless, we confirm that these spurious points are not associated with leakage to the second excited state $\ket{f}$ of the transmon. \cref{sup fig:Second_excited_state} compares the IQ distributions from \cref{sup fig:QND fidelity_extra} with those obtained when the transmon was prepared in state $\ket{f}$. The region in which the spurious points appear does not overlap with the $\ket{f}$ state distribution (shown in green), indicating that they are not caused by leakage to the $\ket{f}$ state. For reference, \cref{sup fig:QND fidelity_extra}(e) shows the IQ distributions for when the transmon is prepared in the state $\ket{g}$, $\ket{e}$ and $\ket{f}$ at the star-marked operating point [see \cref{sup fig:QND fidelity_extra}(b)]. Here spurious points consistent with the switching events are still observed in the distribution. 

These observations indicate that switching between different resonator solutions is a likely limitation of the bifurcating readout. For a well-thermalized system, such switching events are expected to arise from quantum fluctuations \cite{Chen2023}. In future devices, their detrimental impact could be mitigated by choosing parameters such that optimal qubit readout is achieved in a regime where the typical switching time is much longer than the readout integration time.

\section{Contributions to the cross-Kerr shift}
\label{Appendix_cross_kerr_contributions}

In junction readout, the strong nonperturbative cross-Kerr interaction used for readout originates from the $\cos{\hat{\varphi}_t} \cos{\hat{\varphi}_r}$ coupling between the transmon and resonator in the interaction Hamiltonian \cref{eq:int_hamiltonian}, while the unwanted transverse interaction from the $\sin{\hat{\varphi}_t} \sin{\hat{\varphi}_r}$ term is canceled using a compensating $\hat{n}_t \hat{n}_r$ interaction. Nevertheless, as shown in \cref{Fig:flux_response}(c), a small residual contribution from the transverse interactions remains, producing a positive correction to the total cross-Kerr shift. In this section, we show that this residual shift arises from the two-mode squeezing interaction between the transmon and resonator, which becomes enhanced in the vicinity of the balanced condition.

To simplify the discussion, we first neglect the $\cos{\hat{\varphi}_t} \cos{\hat{\varphi}_r}$ interaction. Expanding $\sin{\hat{\varphi}_r}$ to first order in the resonator's phase zero-point fluctuation, the Hamiltonian in \cref{Eq:Hamiltonian} can be rewritten as 
\begin{align}
    \hat{H} = \hat{H}_t + \hat{H}_r + \hat{B} \hat{a}^\dagger + \hat{B}^\dagger \hat{a},
\end{align}
where $\hat{B} = - E_{Jc} \varphi_{\rm{zpf,r}} \sin{\hat{\varphi}_t} + i J n_{\rm{zpf,r}} \hat{n}_t$. In the dispersive regime, a standard Schrieffer–Wolff transformation yields, to leading order, an effective Hamiltonian of the form, 
\begin{align}
    \hat{H} \simeq \hat{H}_t + \hat{H}_r + \sum_j \chi_j \hat{a}^\dagger \hat{a} \ket{j} \bra{j},
\end{align}
where $\ket{j}$ are the eigenstates of the transmon. For simplicity we have absorbed the Lamb shifts into $\hat{H}_t$. 
The dispersive shifts are given by
\begin{align}
    \chi_j = \sum_i \chi_{ij} - \chi_{ji},
\end{align}
with
\begin{align}
    \chi_{ij} = \frac{\vert g_{ij} \vert^2}{\omega_j - \omega_i - \omega_r},
    \label{eqn:disp_shift_individual}
\end{align}
where $g_{ij} = \langle i \vert \hat{B} \vert j \rangle$ and $\omega_i$ ($\omega_j$) is the bare frequency of the $i$th ($j$th) state of the transmon. 
Within the qubit subspace, the relevant dispersive shift can be defined as $\chi_x = (\chi_1 - \chi_0) / 2$. Note that $\chi_x$ captures the dispersive shift arising solely from the transverse $\sin{\hat{\varphi}_t} \sin{\hat{\varphi}_r}$ and $\hat{n}_t \hat{n}_r$ interactions. It does not include the nonperturbative cross-Kerr contribution from the $\cos{\hat{\varphi}_t} \cos{\hat{\varphi}_r}$ interaction.  

For a transmon, the dispersive shift $\chi_x$ can be accurately approximated by considering only the $0 \leftrightarrow 1$ and $1 \leftrightarrow 2$ transitions, yielding 
\begin{align}
    \chi_x \approx \left( \chi_{01} - \frac{1}{2} \chi_{12} \right) + \left( \chi_{10} - \frac{1}{2} \chi_{21} \right). \label{eq:chi_x_approx}
\end{align}
The first two terms arise from the co-rotating contributions which originate from the beam-splitter interaction, proportional to $\hat{a}\hat{b}^\dagger + \hat{a}^\dagger \hat{b}$, between the transmon and resonator, where $\hat{b}$ is the transmon's annihilation operator in the Kerr oscillator approximation. The last two terms arise from the counter-rotating contributions associated with the two-mode squeezing interaction (proportional to $\hat{a}^\dagger \hat{b}^\dagger + \hat{a} \hat{b}$). In conventional dispersive readout, where the transmon couples to the resonator through a charge interaction with strength $J$, one has $\hat{B} = i J \hat{n}_t$, implying $\vert g_{ij} \vert = \vert g_{ji} \vert$. Consequently, the co-rotating terms dominate the dispersive shift $\chi_x$. 

\begin{figure}
    \centering
    \includegraphics[width= 1\columnwidth]{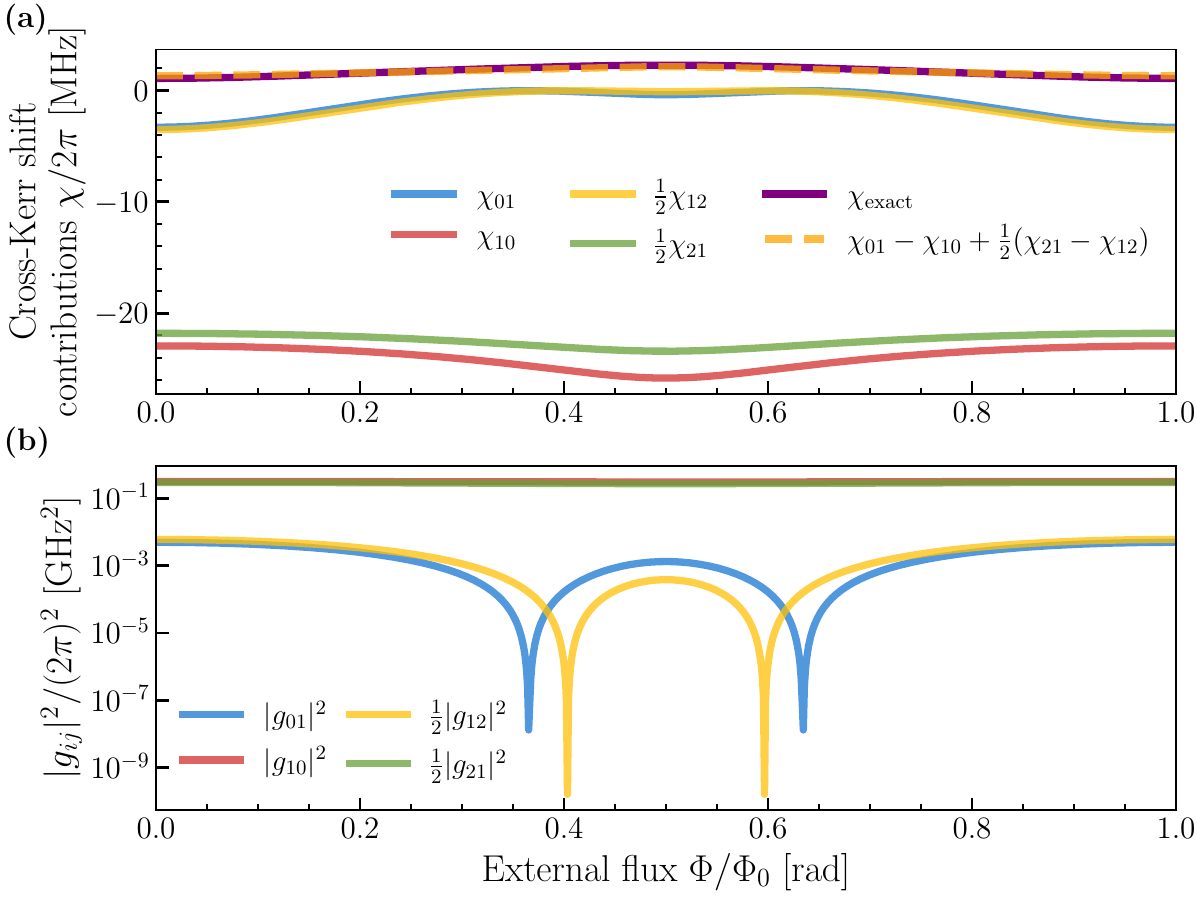}
    \caption{
    \textbf{Cross-Kerr shift contributions.}
    (a) Contributions to the cross-Kerr shift from the relevant dispersive shifts. Here, $\chi_{\rm{exact}}$ is the numerically extracted cross-Kerr shift, while all other values are computed using \cref{eqn:disp_shift_individual} with the fitted Hamiltonian parameters. 
    (b) Numerically extracted coupling coefficient $\vert g_{ij}\vert^2$ where $g_{ij} = \langle i \vert \hat{B} \vert j \rangle$ with $\ket{i}$ and $\ket{j}$ transmon eigenstates.
    }
    \label{sup fig:chi_gij}
\end{figure}

However, for junction readout we have $\vert g_{ij} \vert \neq \vert g_{ji} \vert$. By design, at the balanced point the beam-splitter interaction is suppressed so that $g_{01} = 0$. In contrast, the two-mode squeezing interaction $g_{10}$ is maximized at this same operating point. Equivalently, $g_{10}, g_{21} \gg g_{01}, g_{12}$, which in turn yields $\vert \chi_{10} \vert, \frac{1}{2} \vert \chi_{21} \vert \gg \vert \chi_{01} \vert, \frac{1}{2} \vert \chi_{12} \vert$ as shown in \cref{sup fig:chi_gij}(a) and (b). We find good agreement between the numerically extracted $\chi_{\rm{exact}}$ and the approximate expressions obtained from \cref{eq:chi_x_approx}. Thus, the enhancement of the two-mode squeezing interaction near the balanced condition is responsible for the small contribution of the transverse terms to the overall cross-Kerr shift. 

The preceding discussion, however, does not explain why this small contribution is positive. In fact, when the resonator frequency is above the qubit frequency, as is the case here, the dispersive shift arising from either the $\sin{\hat{\varphi}_t} \sin{\hat{\varphi}_r}$ or $\hat{n}_t \hat{n}_r$ coupling should be negative. Nevertheless, our results suggest that when both interactions are present simultaneously, their combined contribution to the dispersive shift is positive. Remarkably, we find that this net positive contribution originates from the third–order nonlinearity of the $\sin\hat{\varphi}_t$ operator.

To see this, we first note that $\omega_{10} + \omega_r$ and $\omega_{21} + \omega_r$ differ only by the transmon anharmonicity which is small compared to the sum of the qubit and resonator frequencies. Therefore, it is sufficient to compare the matrix elements $g_{10}$ and $g_{21}$ to show that $|\chi_{10} |> |\chi_{21}| / 2$ which leads to an overall positive contribution to $\chi_x$. Expanding the $\sin{\hat{\varphi}_t}$ operator to third order in $\varphi_{\rm{zpf,t}}$, one finds,
\begin{align}
    g_{10} &\approx -E_{Jc} \varphi_{\rm{zpf,r}} (\varphi_{\rm{zpf,t}} - \varphi_{\rm{zpf,t}}^3/2) - J n_{\rm{zpf,r}} n_{\rm{zpf,t}}, \\
    \frac{1}{\sqrt{2}} g_{21} &\approx - E_{Jc} \varphi_{\rm{zpf,r}} (\varphi_{\rm{zpf,t}} - \varphi_{\rm{zpf,t}}^3) - J n_{\rm{zpf,r}} n_{\rm{zpf,t}},
\end{align}
so that $\vert g_{10} \vert > \frac{1}{\sqrt{2}} \vert g_{21} \vert$. Hence, the small positive contribution to the total cross-Kerr shift arises from the third-order nonlinearity of the transmon $\sin{\hat{\varphi}_t}$ operator combined with the two-mode squeezing interaction, which is enhanced near the balanced condition.

\section{Master equation simulation} \label{appendix:purcell_theory}

In this section, we describe how the Purcell-limited lifetimes $T_1^{\rm{pl}}$ shown in \cref{Fig:intrisic_purcell}(a) are computed. For each value of the flux bias, we simulate a \(T_1\) experiment by initializing the qubit in its dressed excited state \(\ket{\overline{1_t, 0_r}}\) and evolving it according to 
\begin{equation}
    \frac{d\hat{\rho}}{dt}= -i[\hat{H},\hat{\rho}] + \kappa \mathcal{D}[\hat{c}_{\mathrm{op}}]\hat{\rho},
\end{equation}
where \(\hat{H}\) is given by \cref{Eq:Hamiltonian}. The transmon and resonator parameters are those extracted from fitting the qubit frequency and anharmonicity versus flux to \cref{Eq:Hamiltonian}, as described in the main text, and we take the resonator decay rate to be $\kappa/2\pi = 10.6$ MHz, as stated previously.

In the above expression, the collapse operator \(\hat{c}_{\mathrm{op}}\) is expressed in the dressed basis as \cite{Nathan2020}
\begin{equation} \label{cop_purcell}
    \hat{c}_{\mathrm{op}} = \sum_{\substack{\lambda',\lambda\\E_{\lambda'}\geq E_\lambda}} \bra{\lambda} \hat{a}^\dagger +\hat{a} \ket{\lambda'} \ket{\lambda}\bra{\lambda'},
\end{equation}
where \(\{\ket{\lambda}\}\) are the eigenstates of the transmon-resonator Hamiltonian \(\hat{H}\) as defined above. Here, we consider only energy-loss processes, reflected in the condition $E_{\lambda'}\geq E_\lambda$
in the sum above.
To extract \(T_1^\mathrm{pl}\), we fit the excited-state population \(\ket{\overline{1_t, 0_r}}\bra{\overline{1_t, 0_r}}\) to a single exponential decay, \(e^{-t/T_1^{\mathrm{pl}}}\).  

To gain further intuition about the different contributions to the Purcell decay, we can express the collapse operator in the Kerr approximation of the transmon. Since we are interested in the single-excitation regime, we can safely neglect higher-order nonlinear terms arising from the coupling junction. Within this approximation, and in the dispersive limit, the collapse operator can be written as
\begin{equation}
\hat{c}_{\mathrm{op}}
= \hat{a}\!+ \!\left(
\frac{\bra{0_t,1_r} \hat{H}_{\mathrm{int}}\ket{1_t,0_r}}{\omega_r - \omega_q}\!
+\!\frac{\bra{0_t,0_r} \hat{H}_{\mathrm{int}}\ket{1_t,1_r}}{\omega_r + \omega_q}
\right)\hat{b},
\end{equation}
where $\hat{a}$ and $\hat{b}$ denote the annihilation operators of the dressed resonator and qubit modes, respectively, and $\hat{H}_{\rm{int}}$ is defined in \cref{eq:int_hamiltonian}. This expression directly leads to \cref{main: T_1 equation} reported in the main text.

\newpage
%


\end{document}